\documentstyle[psfig,times]{mn}

\newcommand{\araa}{ARA\&A}   
\newcommand{\aj}{AJ}         
\newcommand{\aaa}{A\&A}      
\newcommand{\aas}{A\&AS}     
\newcommand{\aar}{A\&AR}     
\newcommand{\apj}{ApJ}       
\newcommand{\apjs}{ApJS}     
\newcommand{\mnras}{MNRAS}   
\newcommand{\nat}{Nat}       
\newcommand{\pasp}{PASP}     
 
\newcommand{\scrm}[1]{\mbox{\scriptsize\rm #1}}

\newcommand{\etal}{et al.\ }

\newcommand{\Mzon}{M$_{\odot}$}

\newcommand{\kms}{\mbox{km s$^{-1}$}}

\newcommand{\Ha}{H$\alpha$}

\newcommand{\Hc}{H$\gamma$}
\newcommand{\Hd}{H$\delta$}

\newcommand{\OI}{[O\,{\sc i}]}

\newcommand{\NII}{[N\,{\sc ii}]}
\newcommand{\SII}{[S\,{\sc ii}]}
\newcommand{\SIII}{[S\,{\sc iii}]}
\newcommand{\HI}{H\,{\sc i}}

\newcommand{\CaII}{Ca\,{\sc ii}}

\newcommand{\multi}{\multicolumn}

\def\picplace#1{\vbox{\hrule\@height 0.4pt\@width\hsize
\hbox to\hsize{\vrule\@width 0.4pt\@height#1\hfil
\vrule\@width 0.4pt\@height#1}\hrule\@height 0.4pt\@width\hsize}}
\def\squareforqed{\hbox{\rlap{$\sqcap$}$\sqcup$}}
\def\sq{\ifmmode\squareforqed\else{\unskip\nobreak\hfil
\penalty50\hskip1em\null\nobreak\hfil\squareforqed
\parfillskip=0pt\finalhyphendemerits=0\endgraf}\fi}

\def\la{\mathrel{\mathchoice {\vcenter{\offinterlineskip\halign{\hfil
$\displaystyle##$\hfil\cr<\cr\sim\cr}}}
{\vcenter{\offinterlineskip\halign{\hfil$\textstyle##$\hfil\cr
<\cr\sim\cr}}}
{\vcenter{\offinterlineskip\halign{\hfil$\scriptstyle##$\hfil\cr
<\cr\sim\cr}}}
{\vcenter{\offinterlineskip\halign{\hfil$\scriptscriptstyle##$\hfil\cr
<\cr\sim\cr}}}}}
\def\ga{\mathrel{\mathchoice {\vcenter{\offinterlineskip\halign{\hfil
$\displaystyle##$\hfil\cr>\cr\sim\cr}}}
{\vcenter{\offinterlineskip\halign{\hfil$\textstyle##$\hfil\cr
>\cr\sim\cr}}}
{\vcenter{\offinterlineskip\halign{\hfil$\scriptstyle##$\hfil\cr
>\cr\sim\cr}}}
{\vcenter{\offinterlineskip\halign{\hfil$\scriptscriptstyle##$\hfil\cr
>\cr\sim\cr}}}}}
\def\degr{\hbox{$^\circ$}}
\def\arcmin{\hbox{$^\prime$}}
\def\arcsec{\hbox{$^{\prime\prime}$}}
\def\utw{\smash{\rlap{\lower5pt\hbox{$\sim$}}}}
\def\udtw{\smash{\rlap{\lower6pt\hbox{$\approx$}}}}

\def\fs{\hbox{$.\!\!^{\rm s}$}}
\def\fdg{\hbox{$.\!\!^\circ$}}
\def\farcm{\hbox{$.\mkern-4mu^\prime$}}
\def\farcs{\hbox{$.\!\!^{\prime\prime}$}}

\def\cor{\mathrel{\mathchoice {\hbox{$\widehat=$}}{\hbox{$\widehat=$}}
{\hbox{$\scriptstyle\hat=$}}
{\hbox{$\scriptscriptstyle\hat=$}}}}

\def\diameter{{\ifmmode\mathchoice
{\ooalign{\hfil\hbox{$\displaystyle/$}\hfil\crcr
{\hbox{$\displaystyle\mathchar"20D$}}}}
{\ooalign{\hfil\hbox{$\textstyle/$}\hfil\crcr
{\hbox{$\textstyle\mathchar"20D$}}}}
{\ooalign{\hfil\hbox{$\scriptstyle/$}\hfil\crcr
{\hbox{$\scriptstyle\mathchar"20D$}}}}
{\ooalign{\hfil\hbox{$\scriptscriptstyle/$}\hfil\crcr
{\hbox{$\scriptscriptstyle\mathchar"20D$}}}}
\else{\ooalign{\hfil/\hfil\crcr\mathhexbox20D}}%
\fi}}

\def\bbbc{{\mathchoice {\setbox0=\hbox{$\displaystyle\rm C$}\hbox{\hbox
to0pt{\kern0.4\wd0\vrule height0.9\ht0\hss}\box0}}
{\setbox0=\hbox{$\textstyle\rm C$}\hbox{\hbox
to0pt{\kern0.4\wd0\vrule height0.9\ht0\hss}\box0}}
{\setbox0=\hbox{$\scriptstyle\rm C$}\hbox{\hbox
to0pt{\kern0.4\wd0\vrule height0.9\ht0\hss}\box0}}
{\setbox0=\hbox{$\scriptscriptstyle\rm C$}\hbox{\hbox
to0pt{\kern0.4\wd0\vrule height0.9\ht0\hss}\box0}}}}
\def\bbbq{{\mathchoice {\setbox0=\hbox{$\displaystyle\rm
Q$}\hbox{\raise
0.15\ht0\hbox to0pt{\kern0.4\wd0\vrule height0.8\ht0\hss}\box0}}
{\setbox0=\hbox{$\textstyle\rm Q$}\hbox{\raise
0.15\ht0\hbox to0pt{\kern0.4\wd0\vrule height0.8\ht0\hss}\box0}}
{\setbox0=\hbox{$\scriptstyle\rm Q$}\hbox{\raise
0.15\ht0\hbox to0pt{\kern0.4\wd0\vrule height0.7\ht0\hss}\box0}}
{\setbox0=\hbox{$\scriptscriptstyle\rm Q$}\hbox{\raise
0.15\ht0\hbox to0pt{\kern0.4\wd0\vrule height0.7\ht0\hss}\box0}}}}
\def\bbbt{{\mathchoice {\setbox0=\hbox{$\displaystyle\rm
T$}\hbox{\hbox to0pt{\kern0.3\wd0\vrule height0.9\ht0\hss}\box0}}
{\setbox0=\hbox{$\textstyle\rm T$}\hbox{\hbox
to0pt{\kern0.3\wd0\vrule height0.9\ht0\hss}\box0}}
{\setbox0=\hbox{$\scriptstyle\rm T$}\hbox{\hbox
to0pt{\kern0.3\wd0\vrule height0.9\ht0\hss}\box0}}
{\setbox0=\hbox{$\scriptscriptstyle\rm T$}\hbox{\hbox
to0pt{\kern0.3\wd0\vrule height0.9\ht0\hss}\box0}}}}
\def\bbbs{{\mathchoice
{\setbox0=\hbox{$\displaystyle     \rm S$}\hbox{\raise0.5\ht0\hbox
to0pt{\kern0.35\wd0\vrule height0.45\ht0\hss}\hbox
to0pt{\kern0.55\wd0\vrule height0.5\ht0\hss}\box0}}
{\setbox0=\hbox{$\textstyle        \rm S$}\hbox{\raise0.5\ht0\hbox
to0pt{\kern0.35\wd0\vrule height0.45\ht0\hss}\hbox
to0pt{\kern0.55\wd0\vrule height0.5\ht0\hss}\box0}}
{\setbox0=\hbox{$\scriptstyle      \rm S$}\hbox{\raise0.5\ht0\hbox
to0pt{\kern0.35\wd0\vrule height0.45\ht0\hss}\raise0.05\ht0\hbox
to0pt{\kern0.5\wd0\vrule height0.45\ht0\hss}\box0}}
{\setbox0=\hbox{$\scriptscriptstyle\rm S$}\hbox{\raise0.5\ht0\hbox
to0pt{\kern0.4\wd0\vrule height0.45\ht0\hss}\raise0.05\ht0\hbox
to0pt{\kern0.55\wd0\vrule height0.45\ht0\hss}\box0}}}}
\def\bbbz{{\mathchoice {\hbox{$\sf\textstyle Z\kern-0.4em Z$}}
{\hbox{$\sf\textstyle Z\kern-0.4em Z$}}
{\hbox{$\sf\scriptstyle Z\kern-0.3em Z$}}
{\hbox{$\sf\scriptscriptstyle Z\kern-0.2em Z$}}}}

\begin{document}

\title[Spectroscopy of star clusters in NGC~1316 (Fornax A)]
{Kinematics, ages, and metallicities of star clusters in 
NGC~1316: \\ A 3 Gyr old merger remnant\thanks{Based on
observations obtained at the European Southern Observatory, La Silla,
Chile (Observing Programme 60.E--0781).}}   

\addtocounter{footnote}{1}

\author[Paul Goudfrooij et al.]{
Paul Goudfrooij,$^{1,2\,}$\thanks{Electronic mail: goudfroo@stsci.edu 
} 
Jennifer Mack,$^{1}$ 
Markus Kissler--Patig,$^{3}$  
Georges Meylan$^{3,1,2}$ 
\newauthor and Dante Minniti\,$^{4}$ 
\\ 
$^1$\,Space Telescope Science Institute, 3700 San Martin Drive,
Baltimore, MD 21218, U.S.A. \\ 
$^2$\,Affiliated to the Astrophysics Division, Space Science
Department, European Space Agency \\ 
$^3$\,European Southern Observatory, Karl-Schwarzschild-Str.~2,
D-85748 Garching, Germany \\ 
$^4$\,Department of Astronomy, P.\ Universidad Cat\'olica, Casilla
306, Santiago 22, Chile \\ 
}

\date{Accepted 2000 October 10; Received 2000 August 10}

\maketitle

\begin{abstract}
We report on multi-object spectroscopy in the red spectral region of 37
candidate star clusters in an $\sim 8 \times 8$ arcmin$^2$ field centered on
the giant early-type radio galaxy NGC~1316 (Fornax~A), the brightest galaxy in
the Fornax cluster.
Out of this sample, 24 targets are found to be genuine star clusters
associated with NGC~1316, 
and 13 targets are 
Galactic foreground stars. For the star cluster sample, we measure a mean
heliocentric velocity $v_{\rm hel} = 1698 \pm 46$ \kms\ and a velocity
dispersion $\sigma = 227 \pm 33$ \kms\ within a galactocentric radius of 24
kpc. 
Partly responsible for the velocity dispersion is a significant rotation in
the star cluster system, with a mean velocity of $\sim 175 \pm 70$ \kms\ along
a position angle of $\sim 6\degr \pm 18\degr$. Using the projected mass
estimator and assuming isotropic orbits, the estimated total mass is 
($6.6 \pm 1.7$) $\times$ 10$^{11}$ \Mzon\ within a radius of 24 kpc. The
mass is uncertain by about a factor of two, depending on the orbital
assumptions. 
The implied $\cal{M}$/$L_B$ ratio is in the range 3\,--\,6. 
Four star clusters in our sample are exceptionally luminous  
($M_V < -12.3$). This means that (1) at least this many clusters in
NGC~1316 are up to an order of magnitude more luminous than the most
luminous star cluster in our Galaxy or M\,31, and (2) that the S/N
ratio of their spectra allows us to measure line strengths with good
accuracy.  
By comparing the measured colours and equivalent widths of \Ha\ and the \CaII\
triplet ($\lambda\lambda$\,8498,\,8542,\,8662 \AA) absorption lines for 
those bright star clusters in our sample with those of single-burst 
population models (the Bruzual \& Charlot 1996 models), we find that
they are coeval with an age of 3.0 $\pm$ 0.5 Gyr. Their metallicities
are found to be solar to within $\pm$ 0.15 dex. 
We discuss the properties of the main body of NGC~1316 and
conclude they are consistent with having hosted a major merger 3 
Gyr ago as well. 
The presence of intermediate-age globular clusters in NGC~1316 shows once
again that globular clusters with near-solar metallicity do form during
galactic mergers, and, moreover, that they can {\it survive\/} disruption
processes taking place during the merger (e.g., dynamical friction, tidal
disruption), as well as evaporation. 
In this respect, NGC~1316 provides a hitherto ``missing'' evolutionary
link between young merger remnants of age $\sim$\,0.5 Gyr such as NGC~3597,
NGC~3921 and NGC~7252 on one side, and older giant ellipticals featuring 
bimodal colour distributions on the other side.  
%

\end{abstract}

\section{Introduction}
\label{s:intro}

\subsection{Extragalactic Globular Cluster Systems}

Recent observations with the {\it Hubble Space Telescope (HST)\/} and
large-field ground-based CCD cameras have caused rapid advances in our
knowledge of the formation and evolution of star cluster systems of 
galaxies. Extragalactic star clusters have in recent years established
themselves as potential tracers of the formation and evolution of
galaxies. The number of detailed photometric studies is increasing steadily,
and these studies reveal interesting connections between star cluster systems
and their host galaxies. 

One important, well-known aspect of star cluster systems among galaxies is
that the number of star clusters per unit galaxy luminosity (named the
specific frequency $S_N$) increases systematically from late-type to
early-type galaxies, being $\sim$\,2--3 times higher in elliptical (E)
than in spiral galaxies of type Sb and later (Harris \& van den Bergh 1981;
Harris 1991). This 
fact has been used as an argument against the scenario for forming ellipticals
through mergers (e.g., van den Bergh 1995). However, 
recent observations with the {\it HST\/} have led to a wealth of discoveries
of young star clusters in merging and starburst galaxies. Indeed, it now seems
that star clusters in general (and globular clusters in particular) may form
preferentially in high-density regions of starbursts (e.g., Meurer et al.\
1995), probably from giant molecular clouds whose collapse is being triggered
by a 100-- to 1000-fold increase in gas pressure due to supernova and shock
heating during starbursts (e.g., Elmegreen \& Efremov 1997). 
Galactic gas-rich mergers are known to produce the most energetic known
starbursts (e.g., Sanders \& Mirabel 1996), to create galaxy remnants that
have surface brightness profiles consistent with those of ``old'' giant
ellipticals (Wright et al.\ 1990), and to create large numbers of young
globular clusters (e.g., Whitmore 1999 and references therein). It is
possible that the higher specific frequency of globular clusters in
ellipticals relative to that in spirals may be accounted for by secondary
populations of globular clusters created during gas-rich mergers.

In this respect, one particularly interesting feature of the star cluster
systems of many giant ellipticals is the presence of bimodal colour
distributions, providing clear evidence for the occurrence of a ``second
event'' in the formation of these systems (e.g., Zepf \& Ashman 1993; Whitmore
et al.\ 1995; Geisler, Lee \& Kim 1996; Kissler--Patig et al.\ 1997; Carlson
et al.\ 1998). While such a bimodal colour distribution was actually
predicted from merger models of E galaxy formation by Ashman \& Zepf
\shortcite{ashzep92}, opinions about the general nature of the ``second
event'' differ among authors (see detailed 
review by Ashman \& Zepf 1998). In particular, the ``second event''
could actually be a series of different events that lead to the formation
of metal-rich globular clusters. It is therefore important to obtain any
additional evidence for (or against) a merger origin of the star cluster
bimodality. One important constraint to {\it any\/} scenario to explain the
bimodality is set by the colour of the ``red'' peak in the colour distribution
among giant ellipticals, which (if interpreted in terms of metallicity)
typically indicates a near-solar metallicity (e.g., Forbes, Brodie \&
Grillmair 1997 and references therein). This metallicity estimate has
since been confirmed by Keck/LRIS {\it spectroscopy\/} of globular
clusters in M\,87 (Cohen, Blakeslee \& Rhyzov 1998) and in NGC 1399
(Kissler--Patig et al.\ 1998). In order to render the scenario in
which the second-generation ``red'' star clusters are formed during
galaxy mergers generally applicable, it is therefore of great interest
to find out just what the metallicities of the young, luminous star
clusters in merger remnants actually are, and whether or not they can
be found in merger remnants at different times after the last (major) merger.  

The current evidence that young star clusters in mergers have approximately
solar metallicities is still sparse. Photometry of these clusters typically
does not provide enough information, due to the age-metallicity degeneracy of
(especially optical) colours and the fact that most merger remnants contain 
extensive dust features with appreciable extinction. Progress has recently 
been made by using combinations of near-infrared and optical colours (Maraston 
et al.~2000), but this method is still in its infancy.  The only dependable
evidence stems from relatively strong metallic features observed in {\it
spectra\/} of one bright cluster in the peculiar elliptical galaxy NGC~1275
(Zepf et al.\ 1995a; Brodie et al.\ 1998) and of two such clusters in NGC~7252
\cite{schsei98}, a prototypical, 0.5\,--\,1 Gyr old remnant of two merged
spiral galaxies (Hibbard \& Mihos 1995; Schweizer 1998). It is clear that the
need for more spectroscopic metallicity estimates for young star clusters in 
(other) mergers is imminent.  

Spectroscopy of star clusters around ellipticals can also provide one with
the {\it kinematics\/} of the cluster system, another essential source of
information for discriminating between different galaxy formation scenarios. 
For example, spectroscopic studies of M\,87 and NGC 1399 revealed a
significantly higher velocity dispersion for the globular clusters than for
the stars (Mould et al.~1990; Brodie and Huchra 1991; Grillmair et al.~1994;
Cohen \& Ryzhov 1997; Minniti et al.\ 1998; Kissler--Patig \& Gebhardt
1998; Kissler-Patig et al.\ 1999).
As both M\,87 and NGC~1399 are central cluster ellipticals, these
authors suggested that the globular clusters were reacting to the
gravitational potential of the galaxy cluster as a whole rather than
that of the galaxy itself. In NGC~5128 (Cen~A), Hui et al.\
\shortcite{hui+95} reported rotation 
in the globular cluster system, though only for the metal--rich clusters which
seemed to rotate along with the well-known dust lane. However, 
the situation in NGC 4472 (Sharples et al.~1998) and M87
(Kissler-Patig \& Gebhardt 1998) seems to be different in that the
metal--poor globular clusters seem to dominate the rotation. 

The present paper describes new spectroscopic observations aimed at deriving
kinematics, ages and metallicities for a suitable sample of star clusters in
the merger remnant NGC~1316.  

\subsection{NGC 1316}

\begin{figure}
\centerline{\psfig{figure=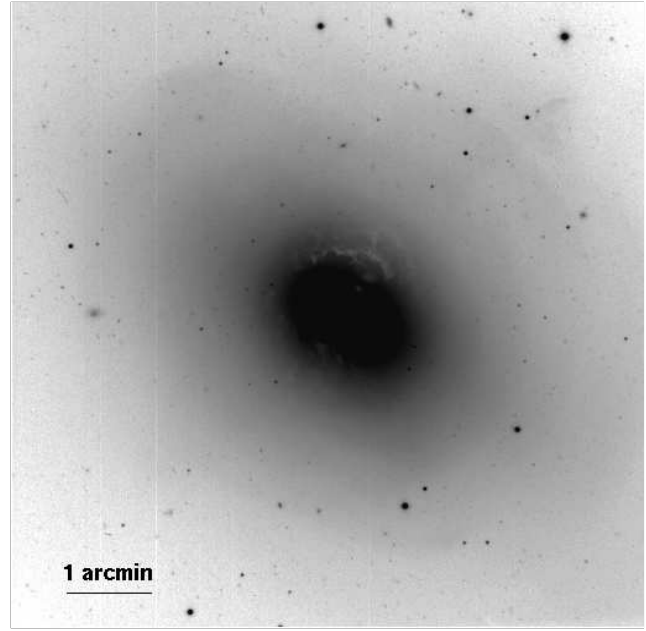,width=8.4cm}}
\caption[]{Grey-scale plot of B-band CCD image of the central 7\farcm5
$\times$ 7\farcm5 of NGC~1316 (taken from
Paper~II). The plate scale is indicated by the bar at the lower left of the
image. North is up and East is to the left.} 
\label{f:fig1}
\end{figure}

The giant early-type galaxy NGC~1316 (= Fornax A = PKS 0320\,$-$\,37 = ESO
357\,$-$\,G022 = ARP 154) is one of the brightest and closest radio galaxies
in the sky. It is located in the outskirts of the Fornax cluster, at a
projected distance of 3\fdg7 from NGC~1399, the central giant elliptical
galaxy. The extensive optical observations of Schweizer (1980) showed
that NGC~1316 is a typical Morgan D-type galaxy with an elliptical
spheroid embedded in an extensive, box-shaped envelope. Several features of
NGC~1316 establish firmly that it is a merger remnant. The outer envelope
includes several non-concentric arcs, tails and loops that are most likely
remnants of tidal perturbations, while the inner part of the spheroid is
characterized by a surprisingly high central surface brightness and small
core radius (and effective radius) for its galaxy luminosity (Schweizer 1981;
Caon, Capaccioli \& D'Onofrio 1994). These characteristics, together with a
velocity dispersion which is significantly lower than that of other elliptical
galaxies of similar luminosity, cause NGC~1316 to lie far off the fundamental
plane of early-type galaxies or the Faber-Jackson relation (e.g., D'Onofrio et
al.\ 1997). Furthermore, the inner spheroid is marked by non-concentric ripples
of 0.1--0.2 mag amplitude above a best-fitting $r^{1/4}$ law (see discussion
in Sect.~\ref{s:dyn}.3).  Figure \ref{f:fig1} shows a $B$-band image of the
inner 7\farcm5\,$\times$\,7\farcm5 of NGC~1316 (taken from Goudfrooij, Alonso
\& Minniti 2000) in which some of the features mentioned above can be found.
All of these features are consistent with NGC~1316 having undergone a
relatively recent merger after which dynamical relaxation has not yet had time
to complete fully.   

\begin{table}
\caption[ ]{Global properties of NGC~1316.}
\label{t:n1316}
\begin{tabular*}{8.4cm}{@{\extracolsep{\fill}}lll@{}} \hline \hline
\multi{3}{c}{~~} \\ [-1.8ex]                                                
 Parameter & Value & Reference \\ [0.5ex] \hline
\multi{3}{c}{~~} \\ [-1.8ex]  
RA~~(J2000.0)     & 03$^{\rm d}$22$\!^{\rm m}$41\fs710 & {\tt *} \\
DEC~(J2000.0)     & $-$37\degr12$'$\,29\farcs41     	 & {\tt *} \\
   Galaxy Type    & (R$'$)SAB(s)0    	   	& RC3 \\
                  & S0$_1$pec 	     	   	& RSA \\
$r_{\scrm{eff}}$  & 109$''$			& CCD94 \\
$\sigma_0$        & 221 km s$^{-1}$ 		& K00 \\
$B_T$             & 9.42 			& RC3 \\
$A_{B,\,{\rm foreground}}$ & 0.00		& BH84 \\
$(B-V)_{\rm eff}$ & 0.93 			& RC3 \\
$(U-B)_{\rm eff}$ & 0.47 			& RC3 \\
$v_{\rm hel}$  & 1760 km s$^{-1}$		& NED \\
Distance	  & 22.9 Mpc 			& {\tt **} \\
$M_{B_T}^0$       & \llap{$-$}22.38		& {\tt ***} \\
\multi{3}{c}{~~} \\ [-1.8ex] \hline 
\multi{3}{c}{~~} \\ [-1.8ex] 
\end{tabular*}
 
\baselineskip=0.97\normalbaselineskip
{\small
\noindent 
{\sl Notes to Table~\ref{t:n1316}.} \\
{\it Parameters:\/~}$r_{\scrm{eff}} \cor$ Effective radius;
$\sigma_0 \cor$ central velocity dispersion; $B_T \cor$ Total B
magnitude; $A_{B,\,{\rm foreground}} \cor$ Absorption in B band
due to ISM in our Galaxy; $(B-V)_{\rm eff}$,
$(U-B)_{\rm eff} \cor\;$Mean colours inside the effective radius;
$v_{\rm hel} \cor\;$Heliocentric velocity;
$M_{B_T}^0 \cor\;$Total absolute B magnitude, corrected for Galactic
absorption. \\ 
{\it References:\/}~RC3 $\cor$ de Vaucouleurs \etal (1991); RSA $\cor$ Sandage
\& Tammann (1987); CCD94 $\cor$ Caon et al.\ (1994); 
K00 $\cor$ Kuntschner (2000); 
BH84 $\cor$ Burstein \& Heiles (1984); 
NED $\cor$ NASA/IPAC Extragalactic Database (http://nedwww.ipac.caltech.edu); 
{\tt *} $\cor$ derived from archival HST/WFPC2 images; 
{\tt **} $\cor$ derived from supernova light curves, see text; 
{\tt ***} $\cor$ derived from $B_T$ and $A_{B,\,\scrm{foreground}}$.
}

\end{table}

As to the adopted distance of NGC~1316, we take advantage of the fact that two
well-observed type Ia supernovae (SNe\,Ia) have occurred in NGC~1316 (SN1980N
and SN1981D). Using the precise distance indicator for SNe\,Ia that utilizes
the tight relation between their light curve shape, luminosity, and colour
(Riess et al.\ 1998; A.\,G.\ Riess, private communication), we arrive at a
distance of 22.9 ($\pm$ 0.5) Mpc for NGC~1316, equivalent to $(m\!-\!M)_0$ =
(31.80 $\pm$ 0.05).  At this distance, 1$''$ corresponds to 111 pc. Note
that this sets NGC 1316 and the sub-group of galaxies surrounding it slightly 
behind the core of the Fornax cluster (for which $(m\!-\!M)_0$ = 31.54 $\pm$
0.14; Ferrarese et al.~2000). Global
galaxy properties of NGC~1316 are listed in Table~\ref{t:n1316}.   

This paper is built up as follows. Section~\ref{s:sample} describes the sample
selection. The observations are described in section~\ref{s:obs}, while
section~\ref{s:redu} deals with the data reduction. The various results are
presented and discussed in sections~\ref{s:res} and \ref{s:disc},
respectively. Finally, section~\ref{s:concl} summarizes the main conclusions
of this study. 

\section{Selection of star cluster candidates}
\label{s:sample}

\begin{figure}
\centerline{\psfig{figure=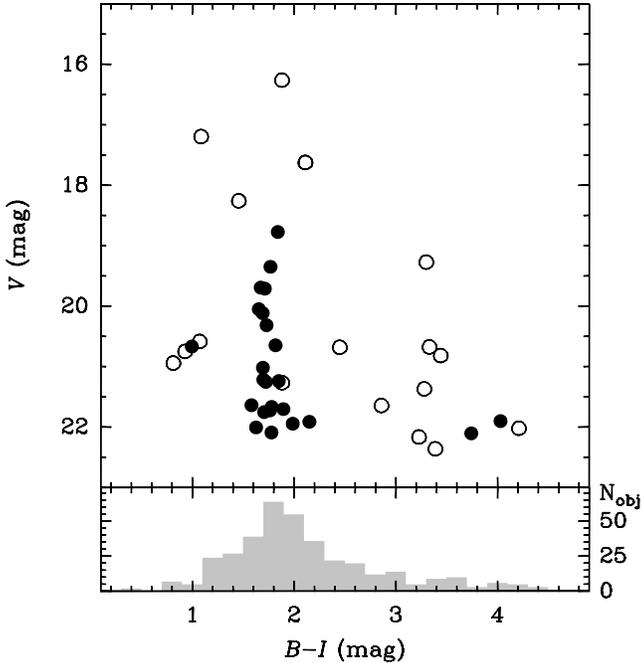,width=8.4cm,angle=-90.}}
\caption[]{Optical colour-magnitude diagram for the sample targets. The open  
circles turned out to be foreground stars after analysis of the spectra
presented in this paper (cf.\ Section~\ref{s:member}), and the filled circles
turned out to be genuine star clusters associated with NGC~1316. A histogram of
the colour distribution of compact objects in the field of NGC~1316 is shown
at the bottom for comparison (taken from Paper~II).}
\label{f:cmd}
\end{figure}

The sample of star cluster candidates was prepared from the target lists of
Shaya et al.\ \shortcite{shay+96} who used the first-generation Planetary
Camera aboard HST ($V$,\,$I$ imaging) and Goudfrooij et al.\ (2000,
hereafter referred to as Paper~II), who used ground-based
($B$,\,$V$,\,$I$,\,$J$,\,$H$,\,$K'$) imaging supplemented with the HST/WFPC2
imaging first published by Grillmair et al.\ \shortcite{gril+99}. 
From these two lists of point-like sources around NGC~1316, a first cut was
made by selecting only targets brighter than $V = 22$ mag, which is the
approximate limiting magnitude to obtain a signal-to-noise (S/N) within an
exposure time of about 4-5 hours on a 4m-class telescope that is high enough 
for reliable velocity measurements. 
Apart from the bright stars that were chosen as pointing checks (see next
Section), a further cut was then made by selecting only targets with $V \ga
18$ (equivalent to $M_V \ga -13.8$ at the assumed distance of NGC~1316), in
order not to miss any possible massive, young or intermediate-age star
clusters. As to the colour selection criterion, we made the following
consideration. The colour distribution of compact objects in NGC~1316 is very 
broad ($0.8 \la B\!-\!I \la 4.5$, cf.\ Paper~II), peaking at about $B\!-\!I =
1.9$. In view of the fact that the central square arcmin of NGC~1316 is full
of dust lanes and patches featuring large extinction values which may redden a
number of star clusters, we considered the full observed range in $B\!-\!I$
for our target selection. A colour-magnitude diagram for the observed targets
is shown in Fig.\ \ref{f:cmd}. The symbols show the different classes of
targets, as resulting from the spectral data shown in this paper. A histogram
of the colour distribution of compact objects in the field of NGC~1316 (from
Paper~II) is added to the figure for comparison.    

\section{Observations}
\label{s:obs}
 
Spectroscopy of the star cluster candidates in NGC~1316 was carried out during
November 15-17, 1997, with the ESO 3.5-m New Technology Telescope (NTT)
equipped with the ESO Multi-Mode Instrument (EMMI) in Multi-Object
Spectroscopy (MOS) mode. The observational setup was identical to the one we
described in detail in Minniti et al.\ \shortcite{minn+98}. Here we give a
somewhat shorter description.   

The detector was a thin, back-illuminated CCD of type Tektronix TK2048EB Grade
2, having 2048$\times$2048 sensitive pixels (of size 24 $\mu$m = 0\farcs268
pixel$^{-1}$). We used grism \#4 (2.8 \AA\ pixel$^{-1}$, blazed at 6500
\AA). The wavelength range is $5500 - 10000$ \AA. 
The MOS slitlets had a width of 1\farcs34, resulting in a wavelength
resolution of 7.5~\AA. We chose grism \#4 in order to include the \CaII\
triplet ($\lambda\lambda$\,8498,\,8542,\,8662 \AA). The \CaII\ triplet index
is an excellent metallicity indicator, being independent of age for stellar 
populations older than about 1 Gyr (e.g., Bica \& Alloin 1987; D\'{\i}az,
Terlevich \& Terlevich 1989; Garc\'{\i}a-Vargas, Moll\'a \& Bressan 1998).    

Once the final target selection was made, it was decided to rotate the
instrument by 90$^{\circ}$ because that made it possible to include all
selected targets in only 2 MOS masks (covering $5' \times 8'$ each), thus
making the most efficient use of telescope time. 
Before punching the slits onto the EMMI MOS plates, the distortion in the
focal plane of the EMMI instrument has to be derived. To this end,
short-exposure R-band images were kindly taken by the ESO NTT team during the
night before our observing run, also using EMMI in the rotated position. The 
($x$,\,$y$) positions of our targets were measured from these images to
produce the MOS masks. In each mask, a few slitlets were punched at positions
corresponding to bright foreground stars in the field of view, which were
used to check the telescope pointing. Furthermore, a few targets were
observed through both masks to enable one to check for any systematic offsets
between radial velocities derived using the different masks. 
One field was centered about 100$''$ north of the centre of NGC~1316, and
observed for a total exposure time of 14,400 s, and the other field (centered
about 90$''$ south of NGC~1316, and including star cluster candidates near the
galaxy centre which, due to the higher galaxy surface brightness, require a
longer exposure time) was exposed for a total of 17,900 s. Individual MOS
exposures were typically of 1800 s duration, interleaved with HeAr
lamp exposures and short imaging
exposures to correct for small pointing errors due to flexure and differential
atmospheric refraction. The latter corrections were always less than
0\farcs3, and the air mass was always kept below 1.3.  
The weather was photometric throughout the observing run, and the seeing
varied from 0\farcs6 to 0\farcs8 as judged from the direct images taken during
the night.

\section{Data reduction and velocity measurements}
\label{s:redu}
The data reduction of the multi-slit spectra was performed in a standard
manner using the IRAF {\sc imred.specred} package, just as described in
Minniti et al.\ \shortcite{minn+98}.  The only noteworthy challenge was a
proper subtraction of the local sky background from the targets that are
close to the galaxy centre, where the galaxy background light is changing
rather rapidly with radius. However, this problem was foreseen:\ We punched
longer slits for those targets near the galaxy centre, enabling us to make
reliable third-order polynomial fits to the galaxy background level as a
function of position along the slit, which usually provided good results. The
only two targets where special care was necessary were targets \#\,112 and
\#\,114 ($\cor$ knots D and E in Schweizer 1980, respectively) (cf.\
Table~\ref{t:posvelmag}), since they are situated 
only 14\farcs5 and 33\farcs9 from the centre, respectively, within a filament
system of ionized gas (cf.\ Schweizer 1980; Mackie \& Fabbiano 1998). This
introduces the additional problem of the presence of radially variable {\it
emission\/} lines.  
This problem was dealt with as follows. We first subtracted the 
emission-line-free background light with the usual third-order polynomial fit
to the background light, but {\it explicitly excluding the wavelength regions
around the emission lines\/} \OI\,$\lambda\lambda6300,6363$,
\NII\,$\lambda\lambda6548,6583$, \SII\,$\lambda\lambda6716,6731$, and
\SIII\,$\lambda\lambda9069,9531$ (redshifted according to the systemic
velocity of NGC~1316) in the fitting procedure. The next (and final) step was
to fit a polynomial to the positions and intensity of those emission lines in
the background-subtracted spectra as a function of position along the slit,
{\it as if they represented a separate system of telluric lines}. The
background in between the emission lines associated with NGC~1316 was median
filtered during the creation of this second ``emission-line sky'' spectrum to
avoid any further degradation of the S/N ratio of the final object
spectrum. This procedure proved successful in the case of target \#\,114,
since its resulting spectrum (cf.\ Fig.\ \ref{f:spectra}) is free of any
residual signs of \NII\,$\lambda6583$ emission. However, the resulting
spectrum of target \#\,112 still contains residual \Ha\ and
\NII\,$\lambda6583$ emission lines, presumably due to its location on a
compact, localized emission-line filament which is not present on the sky
along the slitlet. Hence, target \#\,112 was excluded from the discussion
on line strength indices in Section \ref{s:indices}. 

After wavelength calibration, the spectra were corrected for atmospheric
extinction and flux calibrated using a spectrum of the spectrophotometric
standard star LTT~1020 \cite{ham+94}, which was taken through one of the MOS
slitlets at an airmass similar to those of the NGC 1316 exposures.  

The final spectra were cross correlated with stellar template spectra using
the method of Tonry \& Davis \shortcite{tondav79}, incorporated in the 
{\sc fxcor} task within IRAF. As template spectra we used both the radial
velocity standard HR 1285 (a K0\,{\sc iii} star, taken though a MOS slitlet),
and two bright foreground stars that were already observed as pointing
reference stars. The spectral regions covered by molecular absorption lines
due to the Earth atmosphere were explicitly excluded in the cross
correlations. The zero point of the radial velocities was good to 50
km\,s$^{-1}$ RMS, as judged from the cross correlations of the different 
templates against one another. 

\begin{center}
\begin{table*}
\caption[]{Star clusters and foreground stars: Positions, velocities,
photometry, and classifications.}
\label{t:posvelmag}
\begin{tabular*}{15cm}{@{\extracolsep{\fill}}cccrrrcccl@{}}
    \hline \hline
\multi{5}{c}{~~} \\ [-1.8ex]
\multi{1}{c}{\ } & \multi{1}{c}{RA} & \multi{1}{c}{{\sc Dec}} &
 \multi{1}{c}{{\normalsize $v_{\rm hel}$}} & \multi{1}{c}{$\Delta v$} & 
 \multi{1}{c}{{\normalsize $\epsilon_v$}} & \multi{1}{c}{$V$} & $B\!-\!V$ &
 $B\!-\!I$ & \\ 
\multi{1}{c}{ID} & \multi{1}{c}{(J2000)} & \multi{1}{c}{(J2000)} &
 \multi{1}{c}{(km$\,$s$^{-1}$)} & \multi{1}{c}{(km$\,$s$^{-1}$)} &
 \multi{1}{c}{(km$\,$s$^{-1}$)} & \multi{1}{c}{(mag)} &  \multi{1}{c}{(mag)} &
 \multi{1}{c}{(mag)} & \multi{1}{c}{Classification} \\ 
\multi{1}{c}{(1)} & \multi{1}{c}{(2)} & \multi{1}{c}{(3)} & \multi{1}{c}{(4)}
& \multi{1}{c}{(5)} & \multi{1}{c}{(6)} & \multi{1}{c}{(7)} &
\multi{1}{c}{(8)} & \multi{1}{c}{(9)} & \multi{1}{c}{(10)} \\ [0.5ex] \hline 
\multi{5}{c}{~~} \\ [-0.8ex]
 103 & 03~22~28.47 & $-$37~11~24.5 &  1716 &  15 &  43 & 19.35 & 0.80 & 1.82 &
Globular \\
 104 & 03~22~29.96 & $-$37~11~31.5 &  1895 &  59 &  92 & 21.25 & 0.81 & 1.78 &
Globular \\
 106 & 03~22~32.07 & $-$37~10~36.8 &  2097 &  27 &  88 & 21.66 & 0.83 & 1.83 &
Globular \\
 107 & 03~22~33.32 & $-$37~11~11.9 &  1254 &   3 &  70 & 21.70 & 0.87 & 1.95 &
Globular \\
 110 & 03~22~36.60 & $-$37~10~55.1 &  1919 &   1 &  41 & 20.11 & 0.75 & 1.74 &
Globular \\
\multi{5}{c}{~~} \\ [-1.8ex]
 111 & 03~22~37.90 & $-$37~12~49.8 &  1599 &  59 &  46 & 20.31 & 0.78 & 1.78 &
Globular \\
 112 & 03~22~39.61 & $-$37~12~42.0 &  1271 &   2 &  53 & 19.69 & 0.78 & 1.72 &
Globular \\
 114 & 03~22~42.46 & $-$37~12~40.8 &  1306 &  26 &  44 & 18.77 & 0.86 & 1.89 &
Globular \\ 
 115 & 03~22~43.51 & $-$37~11~26.0 &  1871 &  15 &  74 & 21.91 & 0.75 & 2.20 &
Globular \\
 119 & 03~22~48.90 & $-$37~09~18.3 &  1970 &  60 & 360 & 20.67 & 0.54 & 1.05 &
Globular \\
\multi{5}{c}{~~} \\ [-1.8ex]
 121 & 03~22~50.96 & $-$37~12~27.8 &  1627 & 155 & 240 & 21.75 & 0.90 & 1.76 &
Globular \\
 122 & 03~22~53.14 & $-$37~13~15.8 &  1657 &   4 & 130 & 21.90 & 2.00 & 4.08 &
Globular \\
 123 & 03~22~54.12 & $-$37~10~17.2 &  1966 &   1 &  70 & 20.05 & 0.72 & 1.71 &
Globular \\
 203 & 03~22~31.14 & $-$37~15~25.9 &  1639 &  35 &  68 & 22.09 & 0.83 & 1.83 &
Globular \\
 204 & 03~22~29.35 & $-$37~13~20.9 &  1992 &  19 &  74 & 22.00 & 0.93 & 1.68 &
Globular \\ 
\multi{5}{c}{~~} \\ [-1.8ex]
 205 & 03~22~31.41 & $-$37~12~21.1 &  1569 &  24 &  66 & 20.65 & 0.83 & 1.87 &
Globular \\
 207 & 03~22~33.90 & $-$37~12~31.9 &  1616 &  71 &  63 & 21.72 & 0.83 & 1.82 &
Globular \\
 208 & 03~22~35.28 & $-$37~13~55.0 &  1657 &  34 & 130 & 21.64 & 0.84 & 1.63 &
Globular \\
 210 & 03~22~37.99 & $-$37~13~07.1 &  1451 &   1 &  48 & 19.71 & 0.84 & 1.77 &
Globular \\
 211 & 03~22~39.25 & $-$37~14~24.3 &  1654 &  33 & 123 & 22.10 & 1.76 & 3.79 &
Globular \\
\multi{5}{c}{~~} \\ [-1.8ex]
 212 & 03~22~40.37 & $-$37~13~19.4 &  1752 &   1 &  60 & 21.24 & 0.71 & 1.90 &
Globular \\
 215 & 03~22~46.73 & $-$37~12~11.9 &  1540 &   2 &  61 & 21.02 & 0.79 & 1.75 &
Globular \\
 216 & 03~22~48.01 & $-$37~11~36.6 &  1896 &  36 & 103 & 21.94 & 0.95 & 2.04 &
Globular \\
 217 & 03~22~49.56 & $-$37~11~39.8 &  1840 &   8 &  63 & 21.21 & 0.74 & 1.75 &
Globular \\
\multi{5}{c}{~~} \\ [-1.8ex]
 105 & 03~22~30.64 & $-$37~11~46.0 & $-$25 &  18 &  50 & 21.65 & 1.48 & 2.96 &
Star \\
 108 & 03~22~34.40 & $-$37~10~00.2 & $-$24 &   1 &  21 & 17.20 & 0.41 & 1.19 &
Reference star \\
 109 & 03~22~35.39 & $-$37~09~59.9 &     3 &  36 &  50 & 21.37 & 1.38 & 3.38 &
Star \\
 113 & 03~22~40.73 & $-$37~09~24.1 &   291 & 334 & 107 & 22.17 & 1.52 & 3.33 &
Star \\
 117 & 03~22~46.55 & $-$37~12~27.7 & $-$21 &  46 &  64 & 20.82 & 1.31 & 3.54 &
Star \\
\multi{5}{c}{~~} \\ [-1.8ex]
 118 & 03~22~47.56 & $-$37~13~18.0 &    15 &  12 &  47 & 20.68 & 1.17 & 2.55 &
Star \\ 
 124 & 03~22~55.36 & $-$37~12~32.2 &    20 &   2 &  36 & 20.68 & 1.48 & 3.43 &
Star \\
 125 & 03~22~56.64 & $-$37~11~36.8 &   253 &  13 &  60 & 20.59 & 0.54 & 1.17 &
Star \\
 126 & 03~22~58.23 & $-$37~11~38.2 &     7 &   0 &  30 & 17.63 & 1.04 & 2.21 &
Reference star \\
 202 & 03~22~25.02 & $-$37~12~45.1 &  $-$6 &  15 &  72 & 19.27 & 1.56 & ---  &
Star \\
\multi{5}{c}{~~} \\ [-1.8ex]
 206 & 03~22~32.11 & $-$37~13~07.3 &    35 &   9 & 173 & 22.02 & 1.69 & 4.31 &
Star \\
 209 & 03~22~37.04 & $-$37~14~33.6 &   142 &   1 &  29 & 18.26 & 0.73 & 1.56 &
Reference star \\
 213 & 03~22~42.14 & $-$37~11~14.6 &   430 &   5 &  73 & 21.27 & 0.67 & 1.98 &
Star \\
 214 & 03~22~44.33 & $-$37~12~44.9 &     2 &   4 &  49 & 16.26 & 0.28 & 1.98 &
Reference star \\
 218 & 03~22~51.17 & $-$37~14~20.4 &   165 &  23 &  65 & 20.94 & 0.39 & 0.91 &
Star \\
\multi{5}{c}{~~} \\ [-1.8ex]
 219 & 03~22~52.26 & $-$37~13~33.9 &   126 &  19 & 172 & 22.36 & 1.26 & 3.49 &
Star \\
 220 & 03~22~55.10 & $-$37~15~00.0 &   240 &  57 &  82 & 20.75 & 0.46 & 1.03 &
Star \\
\multi{5}{c}{~~} \\ [-1.8ex] \hline
\multi{5}{c}{~~} \\ [-1.6ex]
\end{tabular*}
\parbox{15cm}{\small \baselineskip=0.95\normalbaselineskip 
{\sl Notes to Table~\ref{t:posvelmag}.} Column (1) contains the target ID;
columns (2) and (3) contain the J2000.0 equatorial coordinates. Units of RA
are hours, minutes, and seconds; units of {\sc Dec} are degrees, arcminutes,
and arcseconds. For comparison, the J2000 coordinates of the nucleus of
NGC~1316 are (RA, DEC) = (03$^{\rm d}$22$^{\rm m}$41\fs710,
$-$37\degr12\arcmin29\farcs41). Column (4) lists the heliocentric radial
velocities; columns (5) and (6) list the velocity uncertainties ($\Delta v$
gives the formal error from the cross-correlation; $\epsilon_v$ gives the
maximum difference of velocities derived using the different
templates). Columns (7), (8) and (9) list the magnitudes and colours,
respectively, and column (10) lists the object classification assigned by
us. ``Reference'' stars designate stars that were used for pointing
checks, cf.\ Section \ref{s:obs}.}  
\end{table*}
\end{center}


\begin{figure*}
\centerline{\psfig{figure=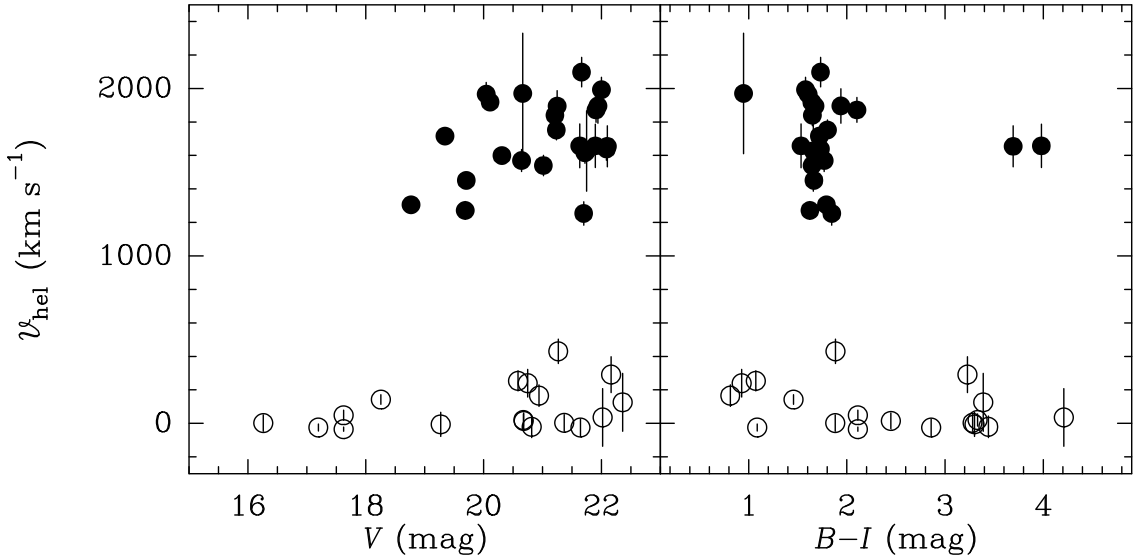,height=7.5cm,angle=-90.}}
\caption[]{Dependence of heliocentric radial velocity on $V$ magnitude {\sl
(left)\/} and $B\!-\!I$ colour {\sl (right)\/} for the foreground stars {\sl
(open circles)\/} and star clusters {\sl (filled circles)\/} of the sample.}
\label{f:stars_gcs}
\end{figure*}

\section{Results}
\label{s:res}

\subsection{Star cluster system membership}
\label{s:member}
 
Table~\ref{t:posvelmag} lists the observed target list of star cluster
candidates in NGC~1316. For each target we give an ID number, the J2000
equatorial coordinates, the heliocentric radial velocity and its
uncertainties, the photometry in $V$, $B\!-\!V$ and $B\!-\!I$ (taken from
Paper~II), and the assigned nature of the target (star\,/\,globular). The
table has been sorted to separate stars and globular clusters for convenience. 

A total of 17 targets are found to be Galactic foreground stars (see 
Table~\ref{t:posvelmag}). Four of these were deliberately chosen as pointing
alignment check (cf.\ Section \ref{s:obs}); the other 13 stars turned out
to be failed star cluster candidates. The remaining 24 targets are 
genuine star clusters associated with NGC~1316. This represents an overall
``success rate'' of 65\%, although this is strongly a function of colour
(e.g., it is 80\% in the range \mbox{1.2 $<B\!-\!I<$ 2.4} in which old,
non-reddened clusters are expected). Figure~\ref{f:stars_gcs} depicts the
dependence of the heliocentric radial velocities on the $V$ magnitudes and
$B\!-\!I$ colours for the foreground stars and the NGC~1316 star
clusters. Note the clear separation in radial velocity between the foreground
stars and the star clusters, and the rather narrow distribution of $B\!-\!I$
colours of the genuine star clusters. Only two of the selected ``red''
targets turned out to be star  clusters. All other targets with $B\!-\!I >
2.1$ are foreground M stars.   

\subsection{Analysis of the star cluster velocity field}
\label{s:dyn}
 
\subsubsection{Velocity distribution of the star clusters}

\begin{figure}
\centerline{\psfig{figure=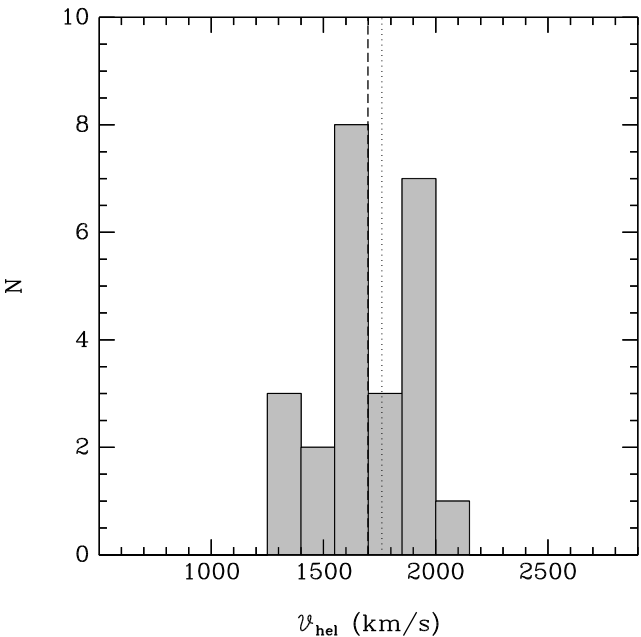,width=8.4cm}}
\caption[]{Heliocentric radial velocity distribution of the 24 star clusters
 in NGC~1316. The dashed line represents the mean velocity of the star
 clusters, and the dotted line represents the systemic velocity 
 of the integrated light of NGC~1316.}
\label{f:velhist}
\end{figure}

Figure~\ref{f:velhist} shows a histogram of the heliocentric velocities of the
star clusters in our sample. The velocity distribution exhibits two peaks:\ 
one on either side of the systematic velocity of the galaxy. 
Fitting a Gaussian to this distribution yields a mean heliocentric velocity
$v_{\rm hel} = 1698 \pm 48$ \kms, similar to the systemic velocity of NGC~1316
(cf.\ Table~\ref{t:n1316}). The formal velocity dispersion of the full sample
is $\sigma = 227 \pm 33$ \kms. However, the double-peaked velocity
distribution signals caution to be taken before assigning a physical meaning
to this value of $\sigma$.  
Our interpretation of the double-peaked velocity distribution is a rotation of
the star cluster system, cf.\ below. 

\subsubsection{Spatial distribution of the velocities}

\begin{figure}
\centerline{\psfig{figure=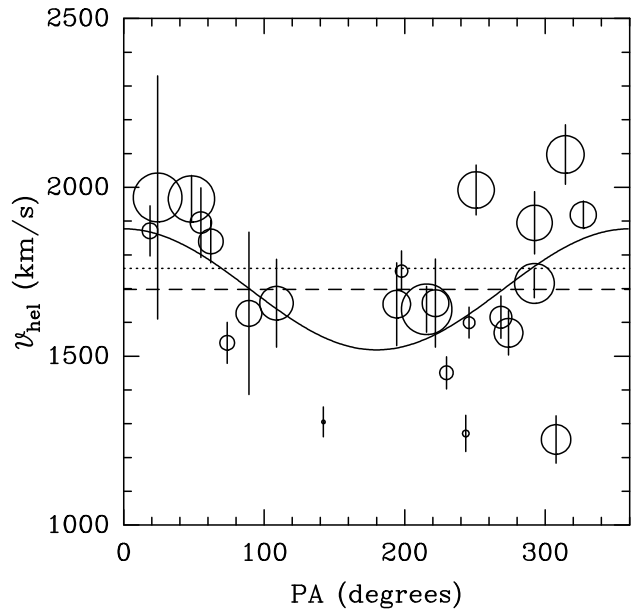,width=8.4cm}}
\caption[]{Radial velocities of the star clusters vs.\ position
 angle measured from the nucleus of NGC~1316 (North through East). The
 symbol size scales linearly with galactocentric radius. The dashed line
 represents the mean radial velocity of the star clusters, the dotted line
 represents the systematic velocity of NGC~1316, and the solid line represents
 a maximum-likelihood cosine fit to the star cluster velocities (see
 Section~\ref{s:dyn} for details).} 
\label{f:gcvel_pa}
\end{figure}

In Figure~\ref{f:gcvel_pa} we depict the relation of the star cluster
velocities with their position angle (hereafter PA; measured from the nucleus
of NGC~1316, North through East), as a function of projected galactocentric
radius. Apart from a few outliers (e.g., the star cluster at PA $\sim$
305\degr, $v_{\rm hel} \sim$ 1250 \kms), the distribution of the velocities in
this plot has a periodic appearance. Interpreting this as rotation of the star
cluster sample, we fitted a flat rotation curve to the heliocentric velocities
using the parametric function 
\[ 
v_{\rm hel} = v_{\rm sys} + v_0 \, \cos (\theta - \theta_0)
\]
where $v_{\rm sys}$ is the systemic velocity, $v_0$ is the rotation velocity,
$\theta$ is the PA of the star cluster, and $\theta_0$ is the PA of the
kinematic major axis of the star cluster system. The three free parameters
were solved for numerically. The uncertainty of the fit is provided by
the standard covariance matrix. The uncertainties of the individual velocities
were also accounted for. However, the uncertainty of the derived rotation
velocity is dominated by the high velocity dispersion of the star cluster
system. The latter was derived after subtracting the best-fitting flat
rotation curve. The fitting procedure was done twice:\ once by leaving
$v_{\rm sys}$ a free parameter, and once by fixing $v_{\rm sys}$ to be the
systemic velocity of the stellar component of NGC~1316, 1760 \kms. The results
of this parametric fitting of the velocity field are listed in
Table~\ref{t:kinfit}.  

\begin{table}
\small 
\caption[]{Results of kinematic fitting of star cluster velocities.}
\label{t:kinfit}
\begin{tabular*}{8.35cm}{@{\extracolsep{\fill}}@{}rcrl@{}} \hline \hline
\multi{3}{c}{~~} \\ [-2.0ex]     
\multi{1}{c}{Rotation} & \multi{1}{c}{Velocity} & \multi{1}{c}{Position} &
 \multi{1}{c}{Mean} \\
\multi{1}{c}{velocity} & \multi{1}{c}{dispersion} & \multi{1}{c}{angle} & 
 \multi{1}{c}{velocity} \\
\multi{1}{c}{(km\,s$^{-1}$)} & (km\,s$^{-1}$) & \multi{1}{c}{(degrees)} &
 \multi{1}{c}{(km\,s$^{-1}$)} \\ 
\multi{1}{c}{(1)} & \multi{1}{c}{(2)} & \multi{1}{c}{(3)} & 
 \multi{1}{c}{(4)} \\ [0.5ex]
\hline
\multi{3}{c}{~~} \\ [-1.8ex]
179 $\pm$ 68 & 202 $\pm$ 33 &  1 $\pm$ 17 & 1698 $\pm$ 48 \\ 
172 $\pm$ 71 & 212 $\pm$ 33 & 10 $\pm$ 19 & 1760$^*$ \\ 
\multi{3}{c}{~~} \\ [-1.8ex] \hline
\multi{3}{c}{~~} \\ [-1.6ex]
\end{tabular*}
\parbox{8.4cm}{
{\sc Note} --- $^*$ Fixed mean velocity to be NGC~1316's systematic velocity.
}
\end{table}

\begin{figure}
\centerline{\psfig{figure=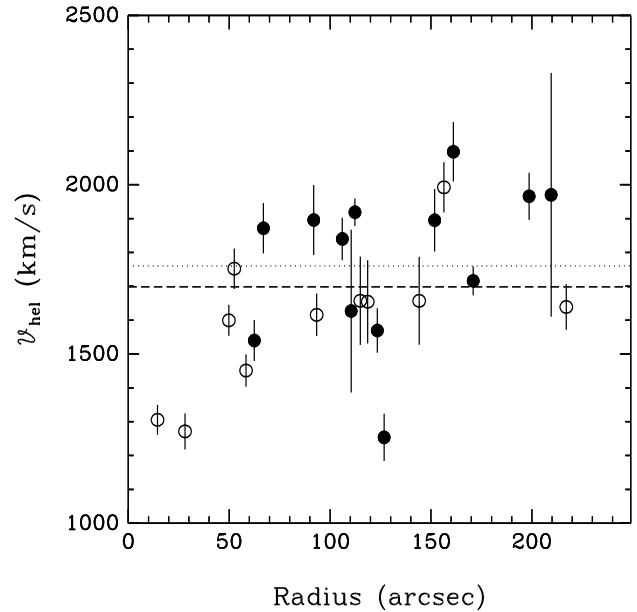,width=8.3cm}}
\caption[]{Radial distribution of the velocities of star clusters in
 NGC~1316. Filled symbols represent star clusters on the north side of the
 apparent major axis of the stellar body NGC~1316; open symbols represent the
 star clusters on the south side of it. Dashed and dotted lines as in
 Figure~\ref{f:gcvel_pa}.} 
\label{f:gcvel_dist}
\end{figure}

No obvious trends of the rotation velocity with galactocentric radius are 
apparent from Figure~\ref{f:gcvel_pa}. This is shown more directly in 
Figure~\ref{f:gcvel_dist}, where we plot radial velocity vs.\ galactocentric
radius for the star clusters. One possibly relevant feature in this plot is
that the two innermost star clusters (within 30$''$ from the centre) both have
a radial velocity that is $\sim$\,450 \kms\ lower than the systematic velocity
of NGC~1316. This difference is (in an absolute sense) $\sim$ 4\,$\sigma$
higher than the rotation velocity derived from the full star cluster
sample. This might indicate either a significant amount of radial motion for
those innermost star clusters, or the presence of differential rotation in the
inner regions.  
 
\subsubsection{Comparison with the kinematics of stars and planetary nebulae}

Arnaboldi et al.\ \shortcite{arna+98} studied the dynamics in the outer
regions of NGC~1316 
using MOS spectra of planetary nebulae in NGC~1316. They find that the
planetary nebulae in NGC~1316 rotate along PA$_{\rm {PN}}$ = 80$^{\circ}
\pm$\,30$^{\circ}$ with a rotation velocity $v_{\rm PN}$ = (140 $\pm$ 60)
\kms\ at a galactocentric radius of 4$'$, and a velocity dispersion 
$\sigma_{\rm PN}$ = (175 $\pm$ 18) \kms. Note that while the rotation velocity
and the velocity dispersion of their planetary nebula sample agree with those
of our star cluster sample to within the combined 1\,$\sigma$ error, the PA of
the kinematic major axis of the planetary nebula sample is different from that
of the star cluster sample by $\sim$\,70\degr\ (although this is formally only
a 2.5\,$\sigma$ result).   

As to the kinematics of the integrated stellar light of NGC~1316, the PA of
the kinematic major axis appears to be consistent with that of the apparent
major axis of the isophotes in the inner $\sim$60$''$, PA$_{\ast}$ $\sim$
58\degr (Arnaboldi et al.\ 1998; Bosma, Smith \& Wellington 1985). Both
Arnaboldi et al.\ and Bosma et al.\ report that the stellar velocity field in
the inner $\sim$\,30$''$ is consistent with NGC~1316 being an oblate,
axisymmetric spheroid flattened by rotation and having isotropic velocity
dispersion.  
However, irregular kinematic features are found at larger distances along both
the major and minor axes: A jump of $\sim$\,30 \kms\ in radial velocity along
the major axis is found at the position of a ripple (Bosma et al.\ 1985), and
similar peculiar velocities and velocity dispersion values are also found
along the minor axis (Arnaboldi et al.\ 1998). These features occur at
positions where prominent shells or ripples are present (as checked on our
direct CCD images). We argue that these features do not affect the global
dynamics significantly, but are associated with the recent merger history of
NGC~1316. Indeed, several authors have argued in the past that shells and
ripples such as those found in NGC~1316 can be formed during a galaxy merger
(e.g., Quinn 1984; Hernquist \& Quinn 1988; Hernquist \& Spergel 1992) or
during a weak interaction (Thompson 1991). The shells and ripples reflect the
various stages of the wrapping of the stars from the smaller galaxy around the
centre of the potential. Peculiar velocities are indeed expected to be
associated with the ripples, and the observed velocity jumps of 30 -- 50 \kms\
are in agreement with the theoretical expectations (cf.\ Quinn 1984).  


\subsubsection{Mass estimates}

\begin{figure*}
\centerline{\psfig{figure=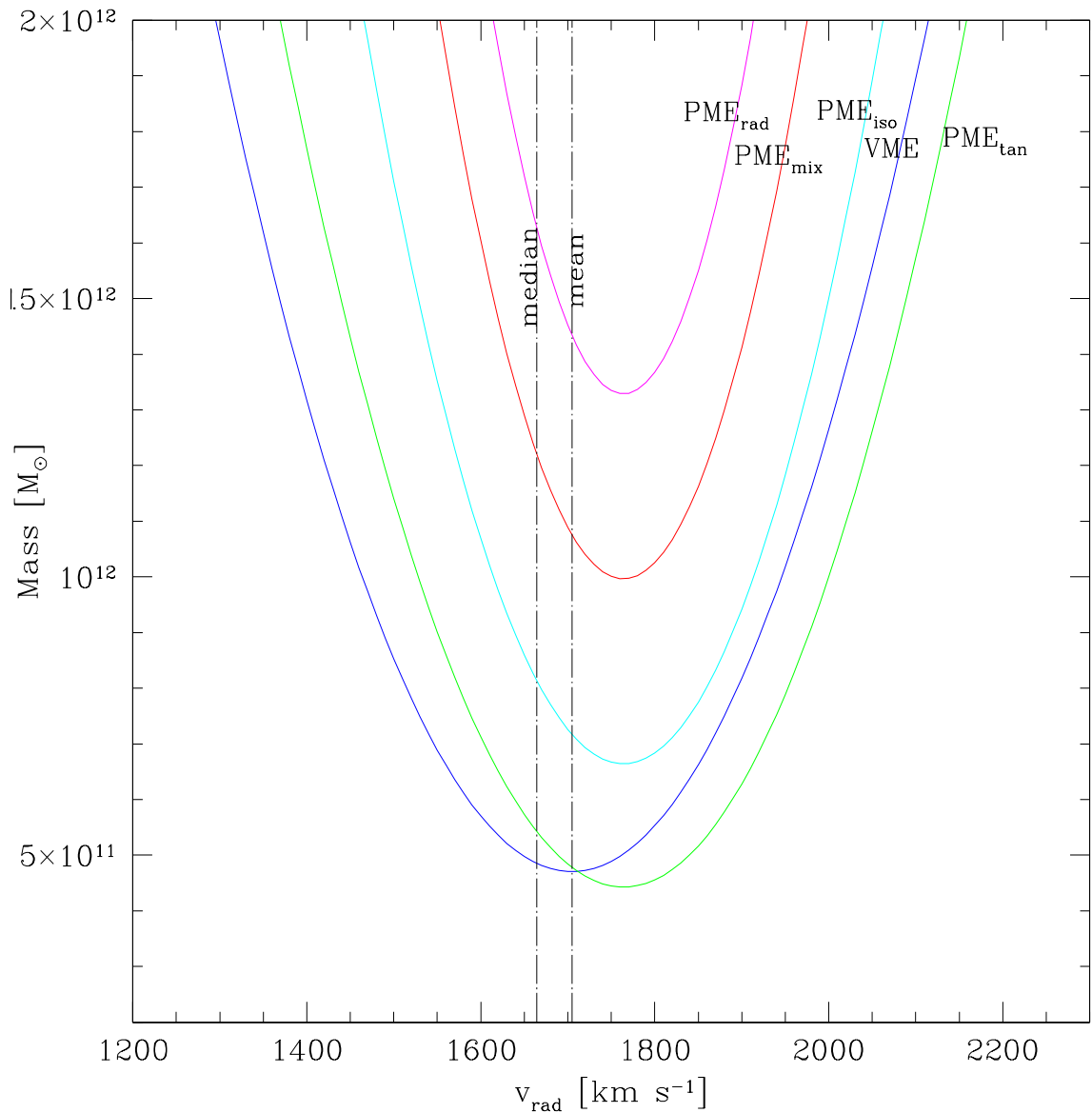,height=15.5cm,width=15.5cm,bbllx=8mm,bblly=57mm,bburx=205mm,bbury=245mm}} 
\caption[]{
Mass estimates for NGC 1316 out to a radius of $r\leq217$\arcsec\ or
$R\leq24$ kpc, as a function of systematic velocity.  
Different orbit characteristics of the globular clusters were assumed for the
mass estimate using the Projected Mass Estimator (PME):\ PME$_{\rm iso}$
assumes isotropic orbits, PME$_{\rm rad}$ assumes radial orbits, 
PME$_{\rm mix}$ assumes mixed orbits, and PME$_{\rm tan}$ assumes tangential  
orbits (for details see Bahcall \& Tremaine 1981). VME shows the results for
the Virial Mass Estimator. The dashed dotted lines show the mean and median
velocity of our globular cluster sample; the systemic velocity of NGC 1316 is 
1760 \kms.}
\label{f:mass_plot}
\end{figure*}

Using the globular clusters in our sample as test particles within the
underlying galaxy potential well, one can estimate a dynamical
mass for the galaxy and derive its mass-to-light ratio for the inner
217\arcsec\ ($\cor$ 24 kpc, the projected distance of our outermost cluster). 
We estimate the mass using all 24 velocities from globular
clusters and applying the virial mass estimator and projected mass
estimator (see, e,g., Bahcall \& Tremaine 1981; Heisler et al.~1985). 
No assumption is made concerning the systemic velocity
which was allowed to vary over a wide range, and masses are computed for 
various assumptions concerning the cluster orbits (radial, tangential, mixed,
or isotropic), following Puzia et al.~(2000). The results are displayed in
Figure \ref{f:mass_plot}. 

Note that the minimum masses are derived for systemic velocities
close to 1760 \kms, the heliocentric velocity of the integrated light of the 
galaxy. The minimum masses vary between $4.4 \times 10^{11}$ M$_\odot$ and
$1.3 \times 10^{12}$ M$_\odot$ depending on the orbital assumptions. Each
measurement has an internal uncertainty of $\sim 30$\% due to the limited
spatial sampling, velocity errors, and sample size, and could rise by $\sim
20$\% by assuming a systematic velocity that is off by 100 \kms. The masses
are also dependent on the assumed distance. For instance, if NGC 1316 would
be at the distance of the central part of the Fornax galaxy cluster as
determined by a compilation of Cepheids, Surface Brightness fluctuations and
Planetary Nebulae luminosity functions in other Fornax cluster galaxies
($(m-M)_0 = 31.30\pm0.04$, Kavelaars et al.~2000), it would reduce the masses
by $\sim 30$\%. The dynamical masses determined using the star clusters as
test particles are consistent with those determined by Arnaboldi et al.\
(1998) using planetary nebulae, taking the difference in the assumed
distances into account. 

\subsubsection{Mass-to-light ratios}

As a working hypothesis to derive mass-to-light ($\cal{M}$/$L$) ratios for the
stellar component of NGC 1316, we will use the mass derived for isotropic
orbits at the measured systemic velocity:\ $(6.6 \pm 1.7) \times 10^{11}$
M$_\odot$.  

Total galaxy luminosities in $B$, $I$, and $H$ were measured in a circular
aperture of 220\arcsec radius (the galactocentric radius of the outermost
star cluster) using our optical and near-infrared surface photometry of the
galaxy (presented in Paper II).  


The results are listed in Table \ref{t:gal_lum}. The table also lists
total luminosities for the integrated light of NGC 1316 after dimming to an
age of 14 Gyr. To estimate 
the latter, we assumed that 
the stars within a 20\arcsec\ radius from
the centre are 2 Gyr old (which is the luminosity-averaged age of the central 2
$\times$ 4\arcsec, spectroscopically measured by Kuntschner 2000). Outside
the 20$''$ radius, we assume that the bulk of the stars are already old,
since the colours and colour gradients in that radial range are consistent
with those of ``normal'' giant elliptical galaxies (Goudfrooij, in
preparation; Schweizer 1980). The actual dimming factors of the composite
population were calculated using the population synthesis models of Bruzual
\& Charlot (1996). The resulting age dimming is only of order 0.1 mag in each
band, both for Salpeter (1955) and Scalo (1986) stellar initial
mass functions (IMFs). While this calculation is only a rough estimate, it
should not be off by any significant amount unless the stellar population of
the {\it whole\/} galaxy would ---unexpectedly--- be only a few Gyr
old. However, we note that the resulting $\cal{M}$/$L$ ratios are rather low
(e.g., $\cal{M}$/$L_B \sim 4.5$), which actually {\it does\/} indicate that
the mean stellar population in NGC~1316 may be younger than usual for giant
elliptical galaxies. 

\begin{table*}[t]
\caption[ ]{Mass-to-light ratios for NGC~1316 within 24 kpc radius.}
\label{t:gal_lum}
\begin{tabular*}{11.2cm}{@{\extracolsep{\fill}}ccccc@{}} \hline \hline
\multi{3}{c}{~~} \\ [-1.8ex]                                                
Filter & Luminosity ($L_\odot$) & $\cal{M}$/$L$ ratio & Corrected
luminosity 
($L_\odot$) & corrected $\cal{M}$/$L$ ratio\\ [0.5ex] \hline
\multi{3}{c}{~~} \\ [-1.8ex]  
$B$ & $1.58 \times 10^{11}$ & 4.2 & $1.43 \times 10^{11}$ & 4.6 \\
$I$ & $2.70 \times 10^{11}$ & 2.4 & $2.46 \times 10^{11}$ & 2.7 \\
$H$ & $7.87 \times 10^{11}$ & 0.8 & $7.18 \times 10^{11}$ & 0.9 \\
\hline
\end{tabular*}

\baselineskip=0.97\normalbaselineskip
{\small
\parbox{11.2cm}{
\smallskip \noindent
{\sl Notes to Table~\ref{t:gal_lum}.} \\
Luminosities are derived assuming M$_{B,\odot}$ = 5.48, M$_{I,\odot}$ = 4.07,
and M$_{H,\odot}$ = 3.39 (Allen 1973; Campins, Rieke \& Lebofsky 1985). 
The listed $\cal{M}$/$L$ ratios have been calculated using isotropic orbits;
it can vary by a factor of 2 depending on the orbit assumptions made}  
}
\end{table*}


\subsection{Star cluster ages and metallicities}
\label{s:indices}

During the early 1990's, a number of different groups have developed
powerful evolutionary population synthesis models employing different stellar
metallicities (Worthey 1994; Vazdekis et al.\ 1996; Bruzual \& Charlot
1996, hereafter BC96; Bressan, Chiosi \& Tantalo 1996; Maraston 1998). 
Hence, the
determination of ages and metallicities of single-burst populations such as
star clusters is now relatively straightforward {\it in principle}. 
However, the determination of ages and metallicities of populations of
non-solar metallicity is hampered by the lack of model spectra of sufficient
spectral resolution for metallicities other than solar. The model spectra of
non-solar metallicity that are used in most spectrophotometric population
synthesis models are based on the Kurucz (1995, private communication)
atmospheres, which have an intrinsic  wavelength resolution of 20 \AA\ in the
optical region. On the other hand, this is not a significant concern for the
spectral indices we chose for deriving ages and metallicities (\Ha\ and
\CaII, see below), since they are well resolved in the model spectra. 

\subsubsection{Line strength measurements}

\begin{figure}
\centerline{\psfig{figure=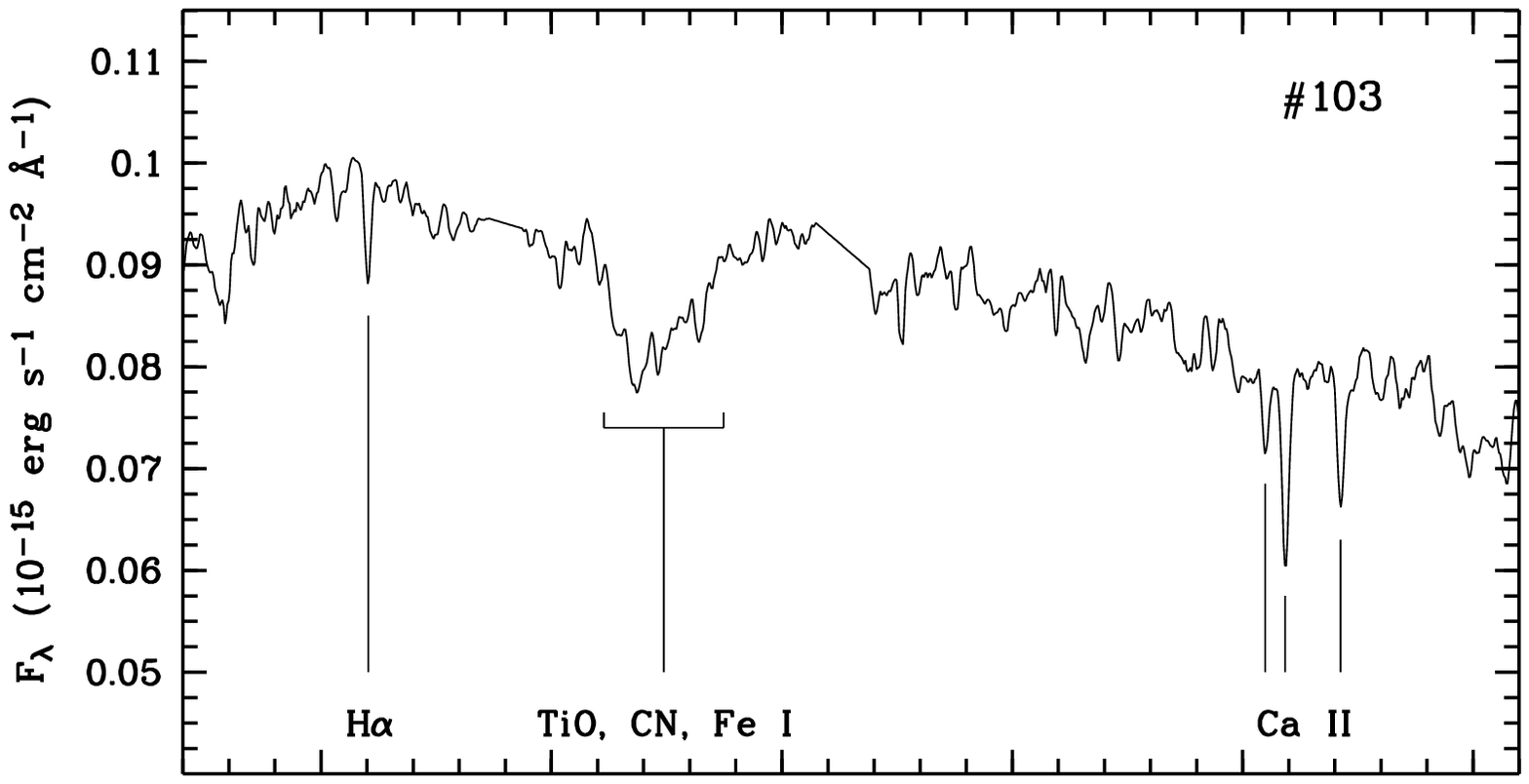,width=8.4cm}}
\vspace*{0.1mm}
\centerline{\psfig{figure=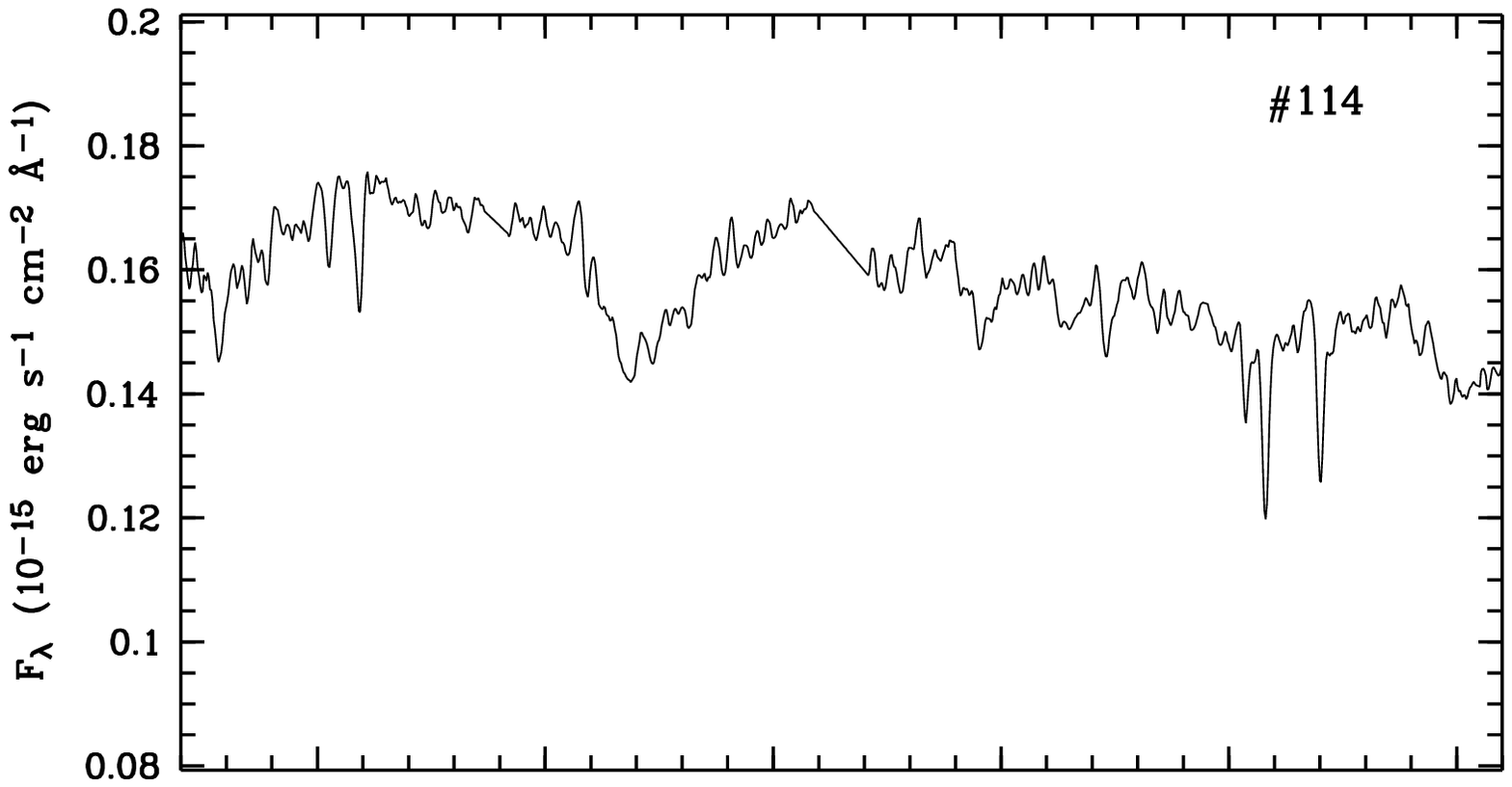,width=8.4cm}}
\vspace*{0.1mm}
\centerline{\psfig{figure=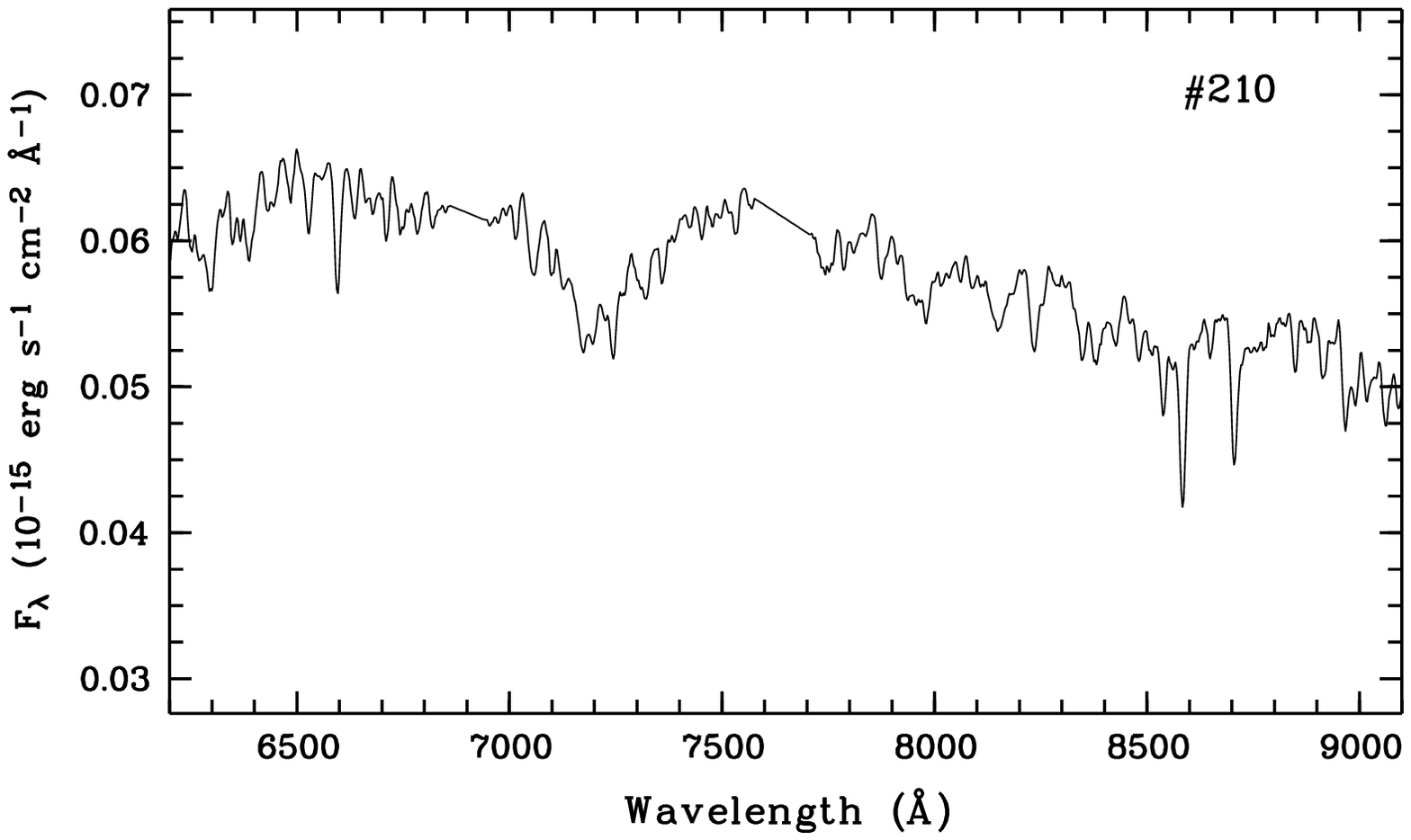,width=8.4cm}}
\caption[]{Flux-calibrated spectra of three bright star clusters in NGC~1316
(ID's 103, 114, and 210). Atmospheric O$_2$ features (6858 -- 6934 \AA, 7580 --
7690 \AA) have been interpolated over. To diminish the influence of noise, the 
spectra have been smoothed with a boxcar kernel of 1 or 2 pixels, depending on
the S/N ratio of their continuum. Main absorption features are identified in
the top panel. Note the relatively strong metal features (compare with
e.g., NGC 1783 and NGC 1978 in Fig.\ 6b of Bica \& Alloin 1987).}
\label{f:spectra}
\end{figure}

\begin{figure*}
\centerline{\psfig{figure=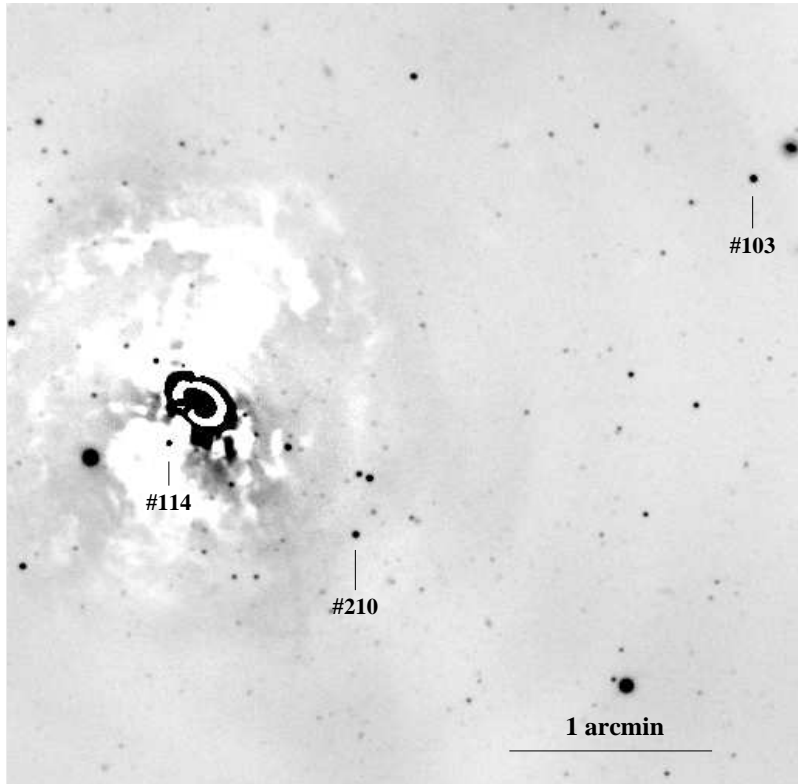,bbllx=2pt,bblly=1pt,bburx=442pt,bbury=440pt,width=10.5cm}}
\caption[]{Negative of $V$-band residual image of NGC~1316 (taken from Paper
II), centered 47$''$ west of the nucleus. White areas (i.e., negative
intensity) indicate galaxy light being absorbed by dust. The black 
elliptical structure (plus white ``ring'') in the central region of
NGC~1316 is a CCD defect caused by saturation. North is up and east is to the
left. The three bright star clusters with ID's 103, 114, and 210 (cf.\ Section
\ref{s:indices}) are indicated, as is the plate scale.}  
\label{f:cluspos}
\end{figure*}

To determine star cluster ages and metallicities, the observed spectra were
convolved with a Gaussian kernel in order to match the wavelength resolution
of the model spectra. Line strengths were then measured from these smoothed
observed spectra, and compared with those measured in an identical manner from
a set of model spectra. To try to achieve a suitable break of the
age--metallicity degeneracy using lines in our wavelength coverage, we
measured \Ha\ (being primarily age sensitive) and the \CaII\ triplet, the
equivalent width (hereafter EW) of which is controlled by the stellar
metallicity for stellar populations older than 1 Gyr (Bica \& Alloin 1987;
D\'{\i}az et al.\ 1989; Garc\'{\i}a-Vargas et al.\ 1998). The line strengths
were measured as described in Goudfrooij \& Emsellem \shortcite{gouems96}. The
errors were quantitatively estimated following Rich \shortcite{rich88}. 
Table~\ref{t:ew} lists the measured equivalent widths for the three bright 
NGC~1316 star clusters with ID's 103, 114, and 210. The spectra of the fainter
clusters had a S/N too low for reliable line strength measurements. 
Figure~\ref{f:spectra} shows the flux-calibrated, extracted spectra of these
star clusters. Their spatial position within NGC~1316 is illustrated in
Figure~\ref{f:cluspos} which depicts the $V$-band ``residual image''
constructed from a $V$-band image after subtraction of an elliptical
isophotal model image (details on this are further discussed in Paper II).  

The adopted passband for EW measurements of \Ha\ was 24~\AA\ wide. As to the
\CaII\ triplet measurements, we adopted the Ca\,T index 
\[ \mbox{EW(Ca\,T)} \equiv \mbox{EW(\CaII\,8542)} + \mbox{EW(\CaII\,8662)} \] 
as defined by D\'{\i}az et al.\ \shortcite{dtt89}. Their definition is well
suited for the resolution of the model spectra.   

\begin{table*}
\small 
\caption[]{Equivalent widths, ages, metallicities, and absolute magnitudes of
 star clusters in NGC~1316} 
\label{t:ew}
\begin{tabular*}{15cm}{@{\extracolsep{\fill}}@{}crrrrccc@{}} \hline \hline
\multi{3}{c}{~~} \\ [-2.0ex]     
\multi{1}{c}{\ } & \multi{1}{c}{\Ha} & \multi{1}{c}{{\CaII}\,8542} &
 \multi{1}{c}{{\CaII}\,8662} & \multi{1}{c}{Ca\,T} & Log Age\rlap{$^{\rm a}$}
 & [$Z$/$Z_{\odot}$]\rlap{$^{\rm a}$}  & $M_V$\rlap{$^{\rm b}$} \\ 
\multi{1}{c}{Cluster ID} & \multi{1}{c}{(\AA)} & \multi{1}{c}{(\AA)} & 
 \multi{1}{c}{(\AA)} &  \multi{1}{c}{(\AA)} & (yr) & (dex) & (mag)
 \\ [0.5ex] 
\hline
\multi{3}{c}{~~} \\ [-1.8ex]
103 & 1.44 $\pm$ 0.08 & 5.17 $\pm$ 0.20 & 3.55 $\pm$ 0.16 & 8.72 $\pm$ 0.26 &
9.45 $\pm$ 0.13 & $+$0.03 $\pm$ 0.04 & $-$12.5 $\pm$ 0.1 \\   
114 & 1.47 $\pm$ 0.06 & 4.64 $\pm$ 0.12 & 3.57 $\pm$ 0.10 & 8.21 $\pm$ 0.16 &
9.46 $\pm$ 0.11 & $-$0.03 $\pm$ 0.03 & $-$13.0 $\pm$ 0.1\\  
210 & 1.50 $\pm$ 0.10 & 4.62 $\pm$ 0.22 & 3.09 $\pm$ 0.18 & 7.71 $\pm$ 0.28 &
9.48 $\pm$ 0.17 & $-$0.13 $\pm$ 0.06 & $-$12.1 $\pm$ 0.1 \\  
\multi{3}{c}{~~} \\ [-1.8ex] \hline
\multi{3}{c}{~~} \\ [-1.6ex]
\end{tabular*}
\parbox{15cm}{
$^{\rm a}$ Ages and [$Z$/$Z_{\odot}$] values estimated from location of
clusters in the \Ha\ vs.\ Ca\,T diagram depicted in Fig.\
\ref{f:CaTHa}. \\
$^{\rm b}$ Derived from $V$ magnitudes in Table \ref{t:posvelmag}, assuming
$m-M$ = 31.8 $\pm$ 0.1. \\
}
\end{table*}

\subsubsection{Age and metallicity estimates}

\begin{figure}
\centerline{\psfig{figure=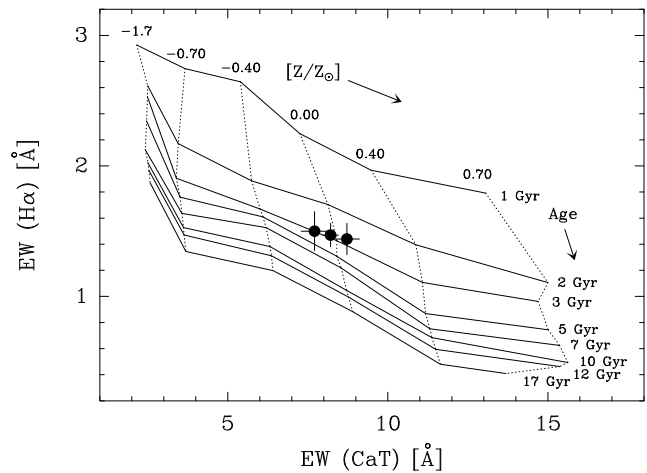,width=8.4cm,angle=-90.}}
\caption[]{EW(\Ha) vs.\ EW(Ca\,T) for clusters 103, 114, and 210 {\it (filled
 circles)}. A grid of isochrones {\it (solid lines)\/} and iso-metallicity
 contours {\it (dotted lines)} based on the Bruzual \& Charlot (1996) models
 is superposed. From this diagram, the cluster metallicities appear to be
 close to solar and the ages $\sim$\,3 Gyr.}
\label{f:CaTHa}
\end{figure}

\begin{figure}
\centerline{\psfig{figure=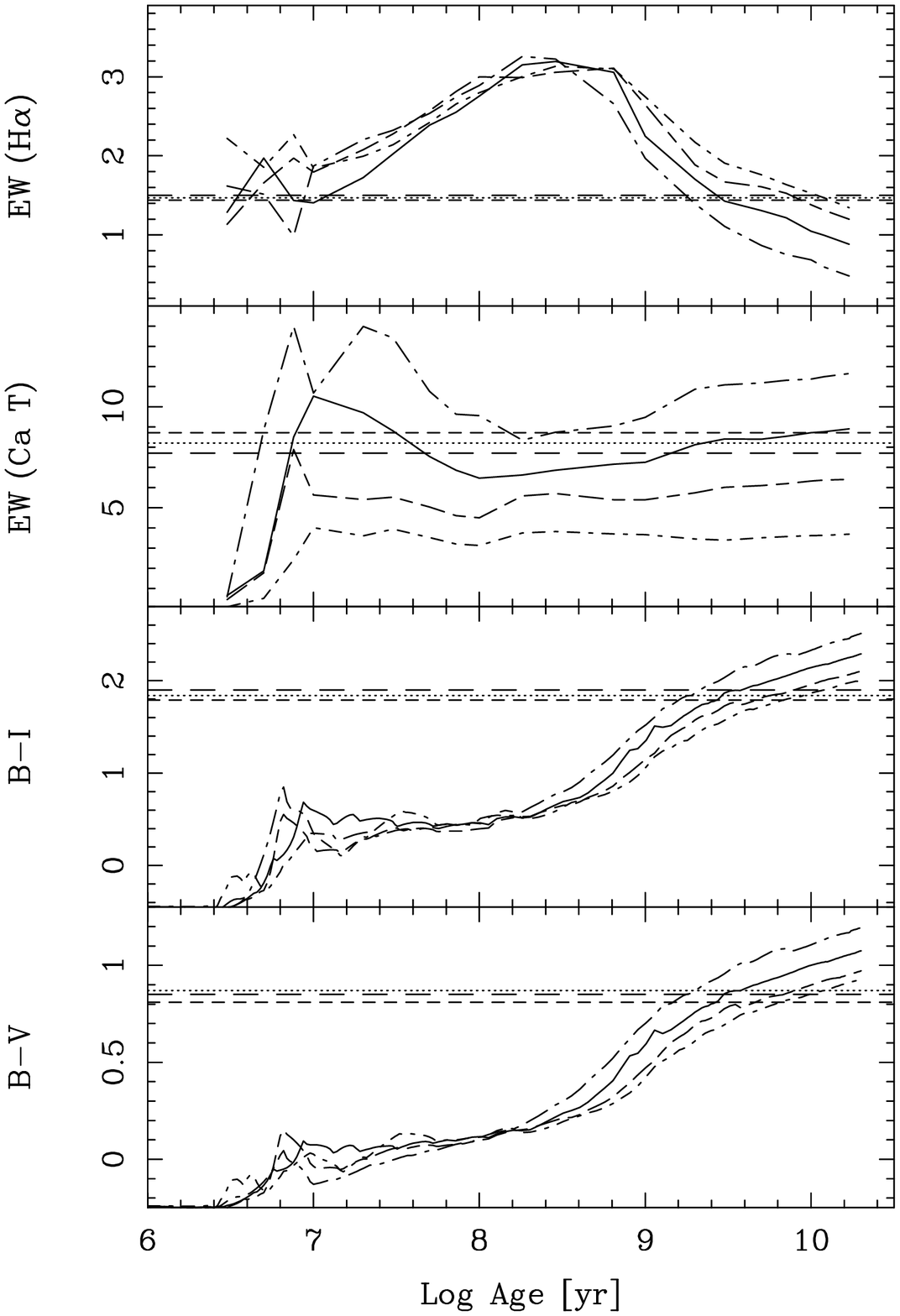,width=8.4cm}}
\caption[]{Time evolution of spectral line indices and colour indices 
 of single-burst stellar populations (Bruzual \& Charlot 1996). Model curves
 are plotted for a Salpeter (1955) IMF and the following metallicities:\ 0.2
 solar {\it (dot-short-dashed lines)}, 0.4 solar {\it (short-long-dashed
 lines)}, 1.0 solar {\it (solid lines)}, and 2.5 solar {\it (dot-long-dashed
 lines)}.  
 The models are compared with observations (horizontal lines) of star clusters
 103 {\it (short-dashed line)}, 114 {\it (dotted line)\/} and 210 {\it
 (long-dashed line)}. Equivalent widths of \Ha\ and Ca\,T are in units of \AA,
 while $B\!-\!V$ and $B\!-\!I$
 are in units of magnitude. The observed line strengths {\it and\/}
 the colour indices of the star clusters are consistent with them having 
 $\sim$\,solar metallicity and an age of $\sim$\,3 Gyr. 
 }
\label{f:agemetal}
\end{figure}

Figures~\ref{f:CaTHa} and \ref{f:agemetal} illustrate the method used to
derive cluster ages and metallicities from the measured equivalent widths. 
Figure~\ref{f:CaTHa} shows a diagram of EW(\Ha) vs. EW(Ca\,T), in which the
drawn lines represent the age--metallicity grid as measured from the model
spectra of different ages and metallicities (BC96). We used the model spectra
of Lejeune, Cuisinier \& Buser \shortcite{leje+97} for our analysis (see BC96
for details). Both Salpeter (1955) and Scalo (1986) stellar initial mass
functions (IMFs) were considered. Since the results were indistinguishable
for our purposes, we only plot the results using the Salpeter IMF in the
diagrams. Note the diagnostic power of the EW(\Ha) vs. EW(Ca\,T) diagram in
breaking the age--metallicity degeneracy, especially for stellar populations
with a (luminosity-weighted) age of 1--3 Gyr. This diagram yields cluster
ages and metallicities encompassing {\it remarkably narrow ranges:\/}
\mbox{2.8 $\la$ Age $\la$ 3.0 Gyr,} and $-$0.13 $\la [Z/Z_{\odot}] \la$ 0.03
(cf.\ Table~\ref{t:ew}).  

These age-- and metallicity estimates are compared to cluster colours in
Figure~\ref{f:agemetal}. The two top panels show the evolution of EW(\Ha) and
EW(Ca\,T) as a function of metallicity, as before. The two other panels in
Figure~\ref{f:agemetal} show the evolution of $B\!-\!V$ and $B\!-\!I$ in the
Kron-Cousins system,  
again using the models of BC96. For comparison with the models,
Figure~\ref{f:agemetal} also shows the measured EWs 
and colours ---marked by horizontal lines--- of the three bright NGC~1316
star clusters. Note that for about solar metallicity (as derived from the
Ca\,T measurements), the star cluster colours indicate Log (Age) $\sim$\,9.45
(i.e., Age $\sim$\,3 Gyr), which is completely consistent with those derived
from EW(\Ha) to within the uncertainties.    

Note that the \Ha\ and Ca\,T equivalent widths of the star clusters
by themselves also allow for an alternative cluster age of $\sim$\,10$^7$ yr
(cf.\ Figure~\ref{f:agemetal}). However, comparison of the colours of the
clusters with those of the BC96 models shows that this alternative solution
would require those clusters to be highly reddened (by \mbox{0.7 $\la
E(B\!-\!V) \la$ 0.8}). Since two of these three clusters (\#103 and \#210) are
located well away from the dust features near the galaxy centre (cf.\ Figure
\ref{f:cluspos}) and since the spectral continuum shape of the three clusters
are extremely similar to one another (cf.\ Figure~\ref{f:spectra}), this
possibility is extremely unlikely (in fact, the similarity between the spectra
and colours of the three clusters argues that cluster \#114 is located in
front of the dust patches onto which it is superposed on the sky). Hence, we
assume in the following that the bright star clusters are $\sim$\,3 Gyr
old. Obtaining additional high-S/N blue-to-visual spectra of these clusters  
(including the higher-order Balmer lines) would certainly enable one to
discriminate conclusively between the two ages described above.  

\section{Discussion}
\label{s:disc}

In this section we discuss various issues regarding the physical nature of the
observed star clusters in NGC~1316, and the relevance of the age and
metallicity of these clusters to {\it (i)\/} observed properties of the main
body of NGC~1316, and {\it (ii)\/} the nature of bimodal colour distributions
of globular cluster systems in ``normal'' giant elliptical galaxies. 

\subsection{Nature of intermediate-age clusters in NGC 1316}

\subsubsection{Cluster luminosities and masses}

Given the distance of NGC~1316 ($(m\!-\!M)_0$ = 31.8), the luminosities of the
three bright star clusters are high ($M_V = -$13.0 for ID\#114, $-$12.5 for
ID\#103, and $-$12.1 for ID\#210, cf.\ Table \ref{t:ew}). For comparison, the
most luminous (by far) globular cluster in our Galaxy, $\omega$\,Cen, has $M_V
= -$10.2 (e.g., Meylan et al.\ 1995), and G1, the most luminous cluster
in M\,31, has $M_V= -$10.85 (Reed et al.\ 1994). 
According to the BC96 models, a 3 Gyr
old cluster with a Salpeter IMF fades by only $\Delta M_V \simeq$ 1.5 mag over
the next 11 Gyr (or by 1.4 mag with a Scalo IMF). Hence, at an age of 14 Gyr,
these three clusters should have absolute magnitudes ($M_V$) of (at least)
$-$11.5, $-$11.0, and $-$10.6, i.e., $\sim$\,2.8, 1.7, and 1.2 times as
luminous as $\omega$\,Cen, respectively. 
%
%
Assuming $\cal{M}$\,/\,$L_V$ = 4.1 as for $\omega$\,Cen \cite{meyl+95}, this
indicates impressive cluster masses of 1.4\,$\times$\,10$^7$ \Mzon\ for \#114, 
8.5\,$\times$\,10$^6$ \Mzon\ for \#103, and 6.0\,$\times$\,10$^6$ \Mzon\ for
\#210, after evolving to an ``old'' age of 14 Gyr. These masses are
similar to that of G1 in M\,31, the most massive globular cluster
known from kinematic measurements ($(1.4 \pm 0.4) \times
10^7$ \Mzon, Meylan et al.\ 1997, 2000). 

\subsubsection{Open or globular clusters\,?} 

Grillmair et al. \shortcite{gril+99} studied the star cluster system of
NGC~1316 using HST/WFPC2 images, whereby they excluded the data within a
$\sim$\,70$''$ galactocentric radius to avoid the dusty areas. 
They found a luminosity function (LF) of the star clusters that does not
resemble the log-normal globular cluster LF observed in almost all globular
cluster systems of ``normal'' (presumed old) elliptical galaxies (see, e.g.,
Ashman \& Zepf 1998). Instead, the LF seemed to increase all the way to the
faintness limits of their photometry. This LF behaviour has been seen in other
merger remnants as well, NGC~4038/4039 (Whitmore \& Schweizer 1995; Whitmore
et al.\ 1999), NGC~3597 \cite{carl+99} and NGC~3256 \cite{zepf+99} being three
well-studied cases. In view of the fact that the open cluster population in
our Galaxy shows an LF whose shape is similar to that of the star clusters in 
NGC~1316, Grillmair et al.\ \shortcite{gril+99} argue that these objects
may be young {\it open\/} clusters rather than globular clusters (an argument
used as well by van den Bergh \shortcite{vdb95} for the case of
NGC~4038/4039). However, the new evidence provided by the much older age of
the bright star clusters in NGC~1316 (with respect to those in the young merger
remnants mentioned above) in combination with the high cluster luminosities
and masses renders the nature of these objects to be genuine {\it globular\/}
clusters beyond any reasonable doubt, at least in NGC~1316. 

As to the apparent disagreement between our conclusion and that of
Grillmair et al.\ \shortcite{gril+99}, we wish to note that only one of the
three clusters for which we have spectroscopic age measurements was covered by
the WFPC2 imaging of Grillmair et al., and that the size distribution of
clusters brighter than $I \sim$ 23.5 is indistinguishable from that in
NGC~1399, an old elliptical galaxy (cf.\ Figure 12 of Grillmair et al.). On
the other hand, we consider it quite likely that the faint end of the observed
cluster system in NGC~1316 may be partly populated by open clusters, in
agreement with Grillmair et al.'s conclusion. 
In fact, these are quite likely candidates of star clusters that will be 
destroyed over the next 10\,--\,12 Gyr as a result of evaporation and/or tidal
shocking (see, e.g., Meylan \& Heggie 1997). Gradual removal of these objects
from the cluster system would conceivably leave behind a log-normal LF as
observed in cluster systems of ``normal'' giant ellipticals. This issue will
be discussed further in Paper~II.  

\subsubsection{Fate of young and intermediate-age clusters}

Extraordinarily luminous, massive, and metal-rich globular clusters such as
the ones found here are now known to be quite common in young merger remnant
galaxies, e.g., NGC 1275 (Zepf et al.\ 1995a, Brodie et al.\ 1998), NGC 7252
(Schweizer \& Seitzer 1998), and NGC 3597 (Carlson et al.\ 1999), all of which
have cluster ages of order 300\,--\,500 Myr. The fate of these massive
clusters has been an issue of some debate. From spectroscopy of luminous
clusters in NGC~1275, Brodie et al.\ (1998) found \Hc\ and \Hd\ EWs that are
somewhat larger than those of any BC96 models that use Salpeter or Scalo
IMFs. They argued that the large Balmer EWs can only be brought into agreement
with the BC96 models by assuming a flatter IMF, which they mimic by
introducing a low-mass cutoff of $\sim$\,2 \Mzon\ and a high-mass cutoff of 
$\sim$\,3 \Mzon\ to the IMF. With such a small or absent
low-mass stellar component, they argue that the young clusters should fade
away in only $\sim$\,1 10$^9$ yr. On the other hand, the $\sim$\,500 Myr young 
clusters in NGC~7252 show Balmer line strengths that are well reproduced by
the BC96 models that use the standard Salpeter or Scalo IMF (Schweizer \&
Seitzer 1998). 
Again, the presence of similarly massive, but $\sim$\,2.5 Gyr older, globular
clusters in NGC~1316 shows that such luminous clusters {\it can\/} survive the
dynamically hot period during the first few Gyr after a galaxy merger. 
We suggest that these high masses be checked through velocity dispersion
measurements 
such as those performed by Ho
\& Filippenko \shortcite{hofil96} for the brightest young cluster in
NGC~1569. One should be able to perform such measurements with 8\,--\,10 m
class telescopes, at least for clusters \#114 and \#103. If the high masses
are confirmed for these 3 Gyr old clusters in NGC~1316, it would provide strong
evidence in support of the view that mergers of (gas-rich) galaxies can form
very massive clusters, with masses similar to or larger than the
largest ones present in our Galaxy or M\,31. 

\subsection{A formation scenario for NGC~1316}

How does the $\sim$\,3 Gyr age of the bright clusters in NGC~1316 relate to
the properties of its main (stellar and gaseous) body? While a multitude of
features of NGC~1316 firmly establish its being a merger remnant (e.g., tidal
tails, shells and ripples, small effective radius; see Introduction), the
complicated structure of NGC~1316 has hampered an accurate reconstruction of
its merger history. 

However, recent numerical simulations of major disk-disk mergers (e.g.,
Hibbard \& Mihos 1995; Barnes \& Hernquist 1996) reveal several features that
are very similar to those in NGC~1316. In particular, we would like to point
out that several properties of NGC~1316 are remarkably consistent with it
being an evolved remnant of a merger event similar to the one which formed
NGC~7252 (0.5\,--\,1 Gyr ago, cf.\ Schweizer \& Seitzer 1998). These
consistencies go far beyond the similarity of the masses and metallicities of
their respective bright globular clusters. 

To set the stage for this comparison, we mention the results of Hibbard et
al.\ (1994)'s optical, X-ray, and \HI\ observations of NGC~7252. \HI\ exists
in large amounts in its tidal tails. However, \HI\ is not detected in its main
stellar body. X-ray emission from hot (10$^6$\,--\,10$^7$ K) gas is detected
from the main stellar body, with a luminosity that is higher than the typical
X-ray luminosity of a spiral galaxy of the same optical luminosity as
NGC~7252. To explain the \HI\ observations, Hibbard \& Mihos (1995) performed
$N$-body simulations of NGC~7252. They successfully modeled the \HI\ velocity 
reversal in the tails as gaseous material from the pre-merger disks that has
reached its orbital turnaround point and is falling back towards the central
regions of the merger remnant. Furthermore, their best-fitting numerical model
for NGC~7252 raised several issues relevant to the current comparison with
NGC~1316, which are discussed below.  
\begin{enumerate}
\item {\it The fate of cold gas in the progenitor disks}.
Extrapolating from the current \HI\ content of NGC~7252's tidal tails,
Hibbard \& Mihos (1995) found that several 10$^9$ \Mzon\ of atomic gas had
already flown inward since the merger took place. As no \HI\ is found in the
inner regions, this atomic gas must have been converted into other
forms. Indeed, hydrodynamical simulations have shown that in the rapidly
changing potential of major mergers, the gas is unable to avoid collisions,
thus dissipating energy and moving inward (e.g., Mihos \& Hernquist 1994). The
high-density clouds within this infalling gas may well be compressed into
molecular form (and subsequently form stars and/or star clusters, depending on
the gas compression achieved (Jog \& Solomon 1992; Elmegreen \& Efremov
1997)), while the low-density gas may shock-heat to the virial temperature
(radiating in the soft X-rays). Indeed, NGC~7252 contains a large amount of
cold molecular gas in its inner regions (Dupraz et al.\ 1990) as well as
extended X-ray-emitting gas (Hibbard et al.\ 1994). Radio continuum data of
the central regions of NGC~7252 (where the molecular gas is found, cf.\
Hibbard et al.\ 1994) indicates a modest star formation rate of 1\,--\,2
\Mzon\ yr$^{-1}$ (Condon 1992). At this rate, the remaining molecular gas
would turn into stars within 2\,--\,4 Gyr (Dupraz et al.\ 1990). During that
time, one would expect the X-ray luminosity to grow as the remaining atomic
and molecular gas contents are shock-heated or evaporated by electrons within
the already existing hot plasma (e.g., Goudfrooij \& Trinchieri 1998). 
Furthermore, if the merger produced a remnant with a deep enough potential
well, the hot plasma would be supplemented by gas donated by mass loss from
evolving giant stars and SNe\,Ia and virialized in the potential well of the
remnant (e.g., Brighenti \& Mathews 1999).  

The observed properties of NGC~1316 fit in remarkably well with this
evolutionary merger scenario: CO observations by Sage \& Galletta (1993)
revealed the presence of $\sim$\,1\,$\times$\,10$^9$ \Mzon\ of molecular
gas. While this is a large amount relative to the molecular gas content of
``normal'' giant elliptical galaxies (see, e.g., Knapp \& Rupen 1996), the
molecular gas is found to be already depleted in the central region, being
concentrated in the dusty regions at $\sim$\,30\,--\,50 arcsec from the centre
instead (Sage \& Galletta 1993).  
In the context of the scenario presented above, this is entirely consistent
with the situation expected a few Gyr after the merger took place. 
Furthermore, a recent analysis of optical spectra of the central 3$''$ of
NGC~1316 reveals a stellar population with a luminosity-weighted age of
$\sim$\,2 Gyr and a metallicity that is as high as other giant elliptical
galaxies of similar luminosity (Kuntschner 2000). This is again entirely
consistent with the scenario of an evolved major merger: The merger occurred
$\sim$\,3\,--\,3.5 Gyr ago (as indicated by the age of the bright globular
clusters) after which stars continued forming within the molecular gas in the
central regions for a few Gyr (the latter process has now apparently largely
ceased). In fact, the unusually small core radius and effective radius of
NGC~1316 (Schweizer 1980; Caon et al.\ 1994) are most probably due to this
star formation episode in the central regions. Finally, the predicted build-up
of hot, X-ray-emitting gas since the merger event has indeed occurred in
NGC~1316: analysis of {\it ROSAT\/} observations have shown the presence of
$\sim$\,10$^9$ \Mzon\ of hot (10$^7$ K) gas, part of which is (still?)
distributed in tails and loops around the galaxy (Kim, Fabbiano \& Mackie
1998).    

\item {\it Tidal features are long-lived}. 
Another important result of the  numerical simulations of NGC~7252 by Hibbard
\& Mihos (1995) was that the tidal tails are not freely expanding. Instead, 
the most tightly bound bases of the tails were found to turn around most
quickly, followed gradually by the material at larger distances from the
centre. The rate of return was found to fall off roughly as $t^{-5/3}$,
implying that tidal features such as those in NGC~7252 stay visible for
several Gyr. In the particular case of NGC~7252 they found that at a time of 4
Gyr after the merger\footnote{Assuming H$_0$ = 75 \kms\ Mpc$^{-1}$}, 
roughly 20\%\ of the current B-band luminosity of the tidal tails remains
outside 4 effective radii from the galaxy centre. 
It is therefore quite plausible that at least some of the large tidal tails
and loops around NGC~1316 still stem from a major merger event similar to the
one that formed NGC~7252, i.e., one during which the 3 Gyr old,
$\sim$\,solar-metallicity globular clusters were formed.  However, we do not
mean to imply that {\it all\/} tails and loops in NGC~1316 necessarily stem
from one merger event. In particular, the tidal tail located $\sim$\,6\farcm5 
southwest of the centre (just outside our CCD field of view, but shown in
Figures 2 and 6 in Schweizer 1980, denoted L$_1$) likely stems from a
significantly more recent interaction with a small, gas-rich galaxy (cf.\
Mackie \& Fabbiano 1998).  
\end{enumerate}

\subsection{Relevance to the nature of bimodal colour distributions in giant
ellipticals} 
\label{s:bimodal}

Assuming that the luminous, metal-rich globular clusters in NGC~1316 were
formed 3 Gyr ago during a major merger involving two spiral galaxies, what
does the expected colour distribution of the resulting globular cluster system
look like? Is it consistent with the observed one? Will it evolve into a
bimodal colour distribution as seen in many ``normal'' giant ellipticals? 

\begin{figure}
\centerline{\psfig{figure=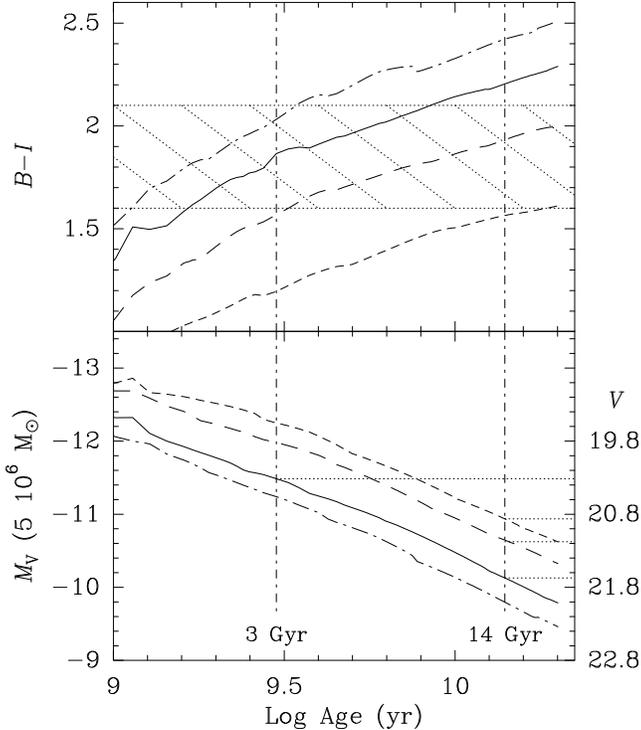,width=8.4cm}}
\caption[]{Time evolution of $B\!-\!I$ colour index {\sl (top
 panel)\/} and absolute magnitude $M_V$ per 5 $\times$ 10$^6$ \Mzon\ {\sl
 (bottom panel)\/} of single-burst stellar population models (Bruzual \&
 Charlot 1996) for ages older than 1 Gyr. Model curves are plotted for a
 Salpeter (1955) IMF and the following metallicities:\ 0.02 solar {\it
 (short-dashed lines)}, 0.2 solar {\it (long-dashed lines)}, 1.0 solar {\it
 (solid lines)}, and 2.5 solar {\it (dot-long-dashed lines)}. Ages of 3 Gyr
 and 14 Gyr are indicated. The region hatched by dotted lines in the top
 panel depicts the $B\!-\!I$ colour interval populated by genuine star
 clusters in NGC~1316 (cf.\ Figure \ref{f:cmd}). 
 The bottom panel shows the predicted evolution of $V$-band magnitudes
 (absolute magnitudes on the left-hand side, apparent magnitudes at
 the distance of NGC~1316 on the right-hand side) for a globular
 cluster having the mass of $\omega$\,Cen. The horizontal dotted lines
 indicate $V$ magnitudes for specific cases, as discussed in
 Section~\ref{s:bimodal}. }
\label{f:cmevol}
\end{figure}

We have seen in Section \ref{s:member} and Figure \ref{f:cmd} that the
observed colour distribution of genuine clusters in NGC~1316 is not bimodal
but spans a rather narrow range around $B\!-\!I$ = 1.85 (cf.\ also Grillmair
et al.\ 1999), 
consistent with the colour predicted by the BC96 models for a 3 Gyr old,
single-burst population of solar metallicity. In contrast, old,
metal-poor, Milky Way halo globular clusters have $B\!-\!I$ colours around 
1.5\,--\,1.6. However, the post-merger cluster
system of NGC~1316 should still contain many of the bulge/halo globular
clusters that were associated with the progenitor galaxies. Is the observed
colour-magnitude diagram consistent with the presence of such ``old''
globulars as well? To address this question, we plot the predicted time
evolution of the $B\!-\!I$ colour and the absolute magnitude $M_V$ for
single-burst populations older than 1 Gyr in Figure~\ref{f:cmevol} (again
using the BC96 models). For 14 Gyr old clusters, the 
observed $B\!-\!I$ colour interval is found to be consistent with a
metallicity range of $0.03 \la Z/Z_\odot \la 0.5$, while the mean colour
$B\!-\!I = 1.85$ would indicate $Z \sim 0.1 \, Z_\odot$. 
While this seems somewhat higher than the median
metallicity of the sample of {\it halo\/} globular clusters in our Galaxy
\mbox{($Z$ = 0.03 $Z_{\odot}$,} cf.\ Harris 1996\footnote{Compiled at
http://www.physics.mcmaster.ca/Globular.html}), it is also somewhat lower
than the median metallicity of the globular cluster system that is associated
with the {\it bulge\/} of our Galaxy ($Z$ = 0.25 $Z_{\odot}$; Minniti 1995).  
This actually suggests that the old clusters in NGC~1316 may be a mixture of
clusters from two progenitor spiral galaxies with different bulge-to-disk
ratios. To assess the reality of this suggestion, we encourage the development
of numerical simulations of interactions of galaxies having bulge and halo
globular cluster systems. The result of such simulations can be expected to
significantly further our understanding of the nature of globular cluster
systems with bimodal colour distributions, and their associated power as
witness of the processes occurring during galaxy formation.  

Finally, the lower panel of Figure~\ref{f:cmevol} depicts the time
evolution of the absolute magnitude $M_V$ (as well as $V$ at the distance of
NGC~1316) for a globular cluster similar to $\omega$\,Cen ($\cal{M}$ = 5
$\times$ 10$^6$ \Mzon, Meylan et al.\ 1995). Taking this to be a typical
brightest globular cluster of any spiral galaxy, one can see that the
``old'', metal-poor clusters are expected to be fainter than $V\sim$
21.0. After the second-generation, solar-metallicity clusters evolve to an
age of 14 Gyr, the BC96 models predict a cluster like $\omega$\,Cen to be
$V\sim$ 21.7 and $B\!-\!I \sim$ 2.2, i.e., those clusters will have $\Delta V
\sim$ 0.7 mag fainter and $\Delta (B\!-\!I) \sim$ 0.3 mag redder (for a given
cluster mass) than the clusters associated with the progenitor
galaxies. These numbers are remarkably consistent with those of typical
globular cluster systems (associated with giant ellipticals) which exhibit
bimodal colour distributions (e.g., Zepf, Ashman \& Geisler 1995; Forbes et
al.\ 1998; Grillmair et al.\ 1999).    

We conclude that the observed colour distribution of clusters in NGC~1316 is 
entirely consistent with that expected from a mixture of {\it (i)\/}
intermediate-age clusters of age $\sim$\,3 Gyr having near-solar metallicities
and {\it (ii)\/} ``old'' ($\sim\,$14 Gyr), metal-poor clusters that were
associated with progenitor galaxies. As such, the colour distribution of this
cluster system is expected to evolve into a bimodal one, having properties
consistent with those of well-studied giant ellipticals. 
The details of the appearance of the final colour distribution of the system
will depend on the ratio of the number of newly formed clusters 
to the number of ``old'' clusters, which is little constrained at this
point. We do know, however, that the massive end of the mass function of the
newly formed clusters is at least as well populated as in the Milky Way and
M\,31 (i.e., as systems with several hundred clusters).  

\section{Main conclusions}
\label{s:concl}

We have obtained spectra for 37 candidate star clusters in the field
of the giant early-type galaxy NGC~1316, an obvious merger remnant featuring
extensive shells, tails, and dusty filaments. Our main conclusions are as
follows.   
\begin{itemize}
\item
The radial velocity measurements reveal that 24 targets are genuine star
clusters, and 13 targets are Galactic foreground stars. 
Almost all (22/24) genuine star clusters fall in the narrow colour
range $1.6 \la B\!-\!I \la 2.1$, i.e., only 2 red targets with $B\!-\!I > 
2.1$ are highly reddened star clusters, the rest are foreground stars.  
\item 
For the star cluster sample, we measure a mean heliocentric velocity
$v_{\rm hel} = 1698 \pm 46$ \kms\ and a velocity dispersion $\sigma = 227 \pm
33$ \kms\ within a galactocentric radius of 24 kpc. Partly responsible for the
velocity dispersion is a significant rotation in the star cluster system, with
a mean velocity of $\sim 175\pm70$ \kms\ along a position angle of $\sim
6\degr \pm 18\degr$. Using the projected mass
estimator and assuming isotropic orbits, the estimated total mass is 
($6.6 \pm 1.7$) $\times$ 10$^{11}$ \Mzon\ within a radius of 24 kpc.
The mass is uncertain by about a factor of two, depending on the orbital
assumptions and adopted distance. The above mass implies a
$\cal{M}$/$L_B$ ratio in the range 3\,--\,6, which is low for
giant elliptical galaxies. This may indicate that the stellar population of
the integrated light of NGC~1316 is, in a luminosity-weighted sense, younger
than usual. 
\item
For three bright star clusters in our sample, the signal-to-noise ratio is
good enough to measure equivalent widths of \Ha\ and the \CaII\ triplet
absorption lines with confidence. Comparing those line strengths as well as
measured broad-band colours with recent single-burst stellar population models
(BC96), the metallicities of those 3 star clusters are found to be solar to
within $\pm$ 0.15 dex, and their most likely age is 3 Gyr.   
\item 
We discuss the properties of the main body of NGC~1316 itself and conclude
they are consistent with having hosted a major merger 3 Gyr ago as well. The
presence of intermediate-age globular clusters in NGC~1316 shows once again
that globular clusters with near-solar metallicity do form during galactic
mergers, and, moreover, that they can {\it survive\/} disruption processes
taking place during the merger (e.g., dynamical friction, tidal disruption),
as well as evaporation.  
In this respect, NGC~1316 provides a hitherto ``missing'' evolutionary link
between the recently studied young merger remnants of $\sim$\,0.5 Gyr age such
as NGC~3921 and NGC~7252 on one side, and ``old'' giant ellipticals featuring
bimodal colour distributions on the other side.   
\end{itemize}

\paragraph*{Acknowledgments.} \ \\ 
PG is grateful to the European Southern Observatory for an invited 
visitorship during which part of this research was carried out, and
for allocating observing time to this project. 
We acknowledge enlightening discussions with Fran\c{c}ois Schweizer,
Claudia Maraston, and Adam Riess.  
We have made use of the NASA/IPAC Extragalactic Database 
(NED) which is operated by the Jet Propulsion Laboratory, Caltech, 
under contract with the National Aeronautics and Space Administration.
JM and PG would like to thank the director of STScI for financial support
through the Director's Discretionary Research Fund. 
The work of DM is partly supported by the Chilean FONDECYT No.\ 01990440.

\end{document}